\documentclass[useAMS,usenatbib]{mn2e}
\usepackage{amsmath}
\usepackage{amssymb}
\usepackage{mathtools}
\usepackage{hyperref}
\usepackage{natbib}
\usepackage{graphicx}
\usepackage{caption}
\usepackage{subcaption}
\usepackage{gensymb}
\usepackage{enumitem}
\DeclareGraphicsExtensions{.pdf,.png,.jpg,.ps}

\title[Vortex avalanches in 2D]
{Simulating pulsar glitches: an $N$-body solver for superfluid vortex motion in two dimensions}
\author[G. Howitt, A. Melatos, B. Haskell]
{G. Howitt$^{1,2}$\thanks{E-mail: ghowitt@student.unimelb.edu.au (GH)},
A. Melatos$^{1,2}$, B. Haskell$^{3}$\\
$^{1}$School of Physics, University of Melbourne, Parkville, Victoria, 3010, Australia\\
$^{2}$OzGrav, Australian Research Council Centre of Excellence for Gravitational Wave Discovery, University of Melbourne, Victoria, 3010, Australia\\
$^{3}$Nicolaus Copernicus Astronomical Center, Polish Academy of Sciences, 
ul. Bartycka 18, 00-716, Warszawa, Poland}
\begin{document}
\date{Accepted XXXX. Received XXXX}

\pagerange{\pageref{firstpage}--\pageref{lastpage}} \pubyear{2014}

\maketitle

\label{firstpage}

\begin{abstract}
\label{abstract}
A rotating superfluid forms an array of quantized vortex lines which determine its angular velocity. 
The spasmodic evolution of the array under the influence of deceleration, dissipation, and pinning forces is thought to be responsible for the phenomenon of pulsar glitches, sudden jumps in the spin frequency of rotating neutron stars. 
We describe and implement an $N$-body method for simulating the motion of up to 5000 vortices in two dimensions and present the results of numerical experiments validating the method, including stability of a vortex ring and dissipative formation of an Abrikosov array.
Vortex avalanches occur routinely in the simulations, when chains of unpinning events are triggered collectively by vortex-vortex repulsion, consistent with previous, smaller-scale studies using the Gross-Pitaevskii equation.
The probability density functions of the avalanche sizes and waiting times are consistent with both exponential and log-normal distributions.
We find weak correlations between glitch sizes and waiting times, consistent with astronomical data and meta-models of pulsar glitch activity as a state-dependent Poisson process or a Brownian stress-accumulation process, and inconsistent with a  threshold-triggered stress-release model with a single, global stress reservoir.
The spatial distribution of the effective stress within the simulation volume is analysed before and after a glitch. 

\end{abstract}


\section{Introduction}
\label{intro}
A superfluid supports rotation through the formation of an array of quantized
vortex lines 
\citep{Onsager1949,Feynman1955}.
In a container whose height greatly exceeds its width, the array is rectilinear, if the angular velocity is less than the critical value, where instabilities set in
\citep{Glaberson1974}. 
Over length-scales much larger than the inter-vortex separation, the fluid mimics solid-body rotation, the motion of individual vortices can be averaged over volume, and key aspects of the flow can be described hydrodynamically, e.g. via a multi-fluid system
\citep{Andersson2006}.
However, there are some macroscopic properties of the flow whose treatment requires tracking the motion of individual vortices, e.g. transport coefficients in a vortex tangle
\citep{Peralta2006,Andersson2007},
and far-from-equilibrium phenomena such as vortex pinning
\citep{Alpar1977,Haskell2016,Drummond2018}.

One system where vortex pinning is important is the interior of a neutron star.
Due to their extreme density ($\sim 10^{14}$ g cm$^{-3}$) and relatively low temperature ($k_{\rm{B}} T \approx 10^6$ eV), neutrons inside a neutron star are believed to condense into a superfluid phase
\citep{Baym1971,Pines1985}.
The superfluid neutrons couple loosely to other components of the star, such as the solid crust and an interpenetrating fluid of charged particles.
As the crust brakes electromagnetically, the superfluid vortices migrate outwards.
However, nuclear lattice sites and magnetic flux tubes `pin' the vortices by providing an attractive force, which resists the Magnus force which brings the neutrons and the crust to co-rotation. 
Hence the neutrons lag the crust.
As the lag grows, so does the Magnus force, until the vortices unpin.
If enough vortices unpin, the resulting back-reaction on the crust is observable as an instantaneous increase in the frequency of the pulsar, i.e. a pulsar glitch
\citep{Anderson1975,Haskell2012}.

Another application of vortex pinning with close connections to pulsar glitches is laboratory experiments on magnetic flux tube avalanches in type II superconductors subjected to changing magnetic fields.
In these systems, magnetic flux is distributed as a triangular array of quantized flux tubes
\citep{Abrikosov1957,Bean1964,Fetter1966}.
As the field ramps down, flux tubes are expelled from the superconductor. 
It is observed that pinning of the flux tubes causes the expulsion to occur in bursts involving up to thousands of flux tubes
\citep{Field1995, Altshuler2004}.

Gross-Pitaevskii simulations of small, idealized systems containing $\sim 10^2$ vortices and $\sim 10^4$ pinning sites demonstrate that unpinning occurs collectively 
\citep{Warszawski2011,Warszawski2012,Warszawski2013,Melatos2015}.
Analytic studies have found that collective unpinning is sensitive to the strength of pinning and to stellar parameters such as mass and temperature, which affect how far vortices can move before re-pinning 
\citep{Haskell2016}.
One-dimensional hydrodynamic simulations reveal that accumulation of vorticity in one region of a neutron star can lead to glitch-like travelling waves which reduce differential rotation
\citep{Khomenko2018}.
The latter simulations rely on assumptions of how pinned vortices are distributed within a neutron star which will be improved by a better understanding of the microscopic pinning dynamics, an important motivation for this paper.
Collective unpinning is hard to study theoretically in the many-vortex regime, due to the computational expense of scaling up Gross-Pitaevskii simulations (neutron stars have $\sim 10^{18}$ vortices and $\sim 10^{50}$ pinning sites, for example).
While existing simulations show clear evidence of unpinning knock-on through short-range interactions and long-range acoustic processes
\citep{Warszawski2012},
it is difficult to reliably determine the many-vortex size and waiting time probability distribution functions (PDFs) from simulations of small systems.
More broadly, it is unclear how the knock-on behaviour scales up to larger systems. 
Similar comments apply to experiments with type II superconductors.

The `vortex avalanche' model resembles systems exhibiting self-organized criticality (SOC)
\citep{Bassler1998,Melatos2008},
a paradigm that has found applications in numerous fields of study;
see \citet{Watkins2016} for a review.
Power-law size and exponential waiting-time PDFs are characteristic of SOC. 
They are observed in the glitch histories of two pulsars, PSR J0534+2200 and PSR J1740$-$3015
\citep{Howitt2018}.
Two other pulsars, PSR J0537$-$6910 and PSR J0835$-$4510 exhibit quasiperiodicity in their glitch size and waiting time PDFs, which is also predicted for fast-driven SOC systems
\citep{Jensen1998}.
In experimental studies on superconducting flux tube avalanches, the PDF of burst sizes is a power law, consistent with the predictions of SOC
\citep{Field1995}.

In this methods paper, we describe a two-dimensional, $N$-body vortex filament code that can handle $N \sim 10^4$ vortices, given reasonable computational resources.
In section \ref{sec:vortex_dynamics} we describe the mathematical framework of the solver.
In section \ref{sec:methods} we describe the algorithm and its implementation.
In section \ref{sec:tests} we validate the numerical method through a systematic set of numerical tests.
In section \ref{sec:results} we illustrate the astrophysical applications of the solver by performing numerical experiments
on a pinned, decelerating superfluid to produce SOC-like vortex avalanches and investigate them in the context of pulsar glitches.

\section{Vortex dynamics}
\label{sec:vortex_dynamics}

We study the motion of a system of $N$ point-like rotational vortices in a two-dimensional geometry with cylindrical symmetry.
On scales larger than the inter-vortex separation, the quantum mechanical structure of the vortices can be ignored, and the vortices move according to classical hydrodynamics.
In the absence of external influences such as lift forces and pinning, the velocity $d\mathbf{x}/dt$ of a quantized vortex at position $\mathbf{x}(t)$ is the same as the bulk fluid velocity at $\mathbf{x}(t)$ induced by all the other vortices.
(In practice, we calculate the bulk velocity from the vorticity; see section \ref{subsec:vortex_vortex}).
The convective motion is supplemented by other effects, such as the interaction of vortices with boundaries, impurities, and other, viscous fluid components.

\subsection{Equations of motion}
\label{subsec:equations_of_motion}

In a reference frame that co-rotates with the container, the position of a vortex in Cartesian coordinates, $(x_i,$ $y_i)$, evolves according to
\begin{equation}
\frac{d}{dt} 
\begin{pmatrix}
x_i \\
y_i
\end{pmatrix}
= \mathcal{R}_\phi
\begin{pmatrix}
v_{i,x} \\
v_{i,y}
\end{pmatrix} \, ,
\label{eq:total velocity}
\end{equation}
with
\begin{equation}
v_{i,x} = - \sum_{j \neq i} \frac{\kappa y_{ij}}{r_{ij}^2} 
+ \sum_{j=1}^N \frac{\kappa y_{ij,\rm{image}}}{r_{ij,\rm{image}}^2}
+ \omega y_i
- \sum_k \frac{\partial V(\textbf{x}_i-\textbf{x}_k)}{\partial y_i} 
\label{eq:x velocity}
\end{equation}
\begin{equation}
v_{i,y} = \sum_{j \neq i} \frac{\kappa x_{ij}}{r_{ij}^2} 
- \sum_{j=1}^N \frac{\kappa x_{ij,\rm{image}}}{r_{ij,\rm{image}}^2}
- \omega x_i
+ \sum_k \frac{\partial V(\textbf{x}_i - \textbf{x}_k)}{\partial x_i} \, .
\label{eq:y velocity}
\end{equation}
In \eqref{eq:x velocity} and \eqref{eq:y velocity}, we define 
$\textbf{x}_{ij} = \textbf{x}_i - \textbf{x}_j = (x_{ij},y_{ij})$ to be the displacement between vortices at $\textbf{x}_i$ and $\textbf{x}_j$, with $r_{ij} = \vert \textbf{x}_{ij} \vert$.
Similarly,
$\textbf{x}_{ij, \rm{image}} = \textbf{x}_i - \textbf{x}_{j, \rm{image}} = (x_{ij,\rm{image}},y_{ij,\rm{image}})$ is the displacement between a vortex at $\textbf{x}_i$ and the image vortex of a vortex at $\textbf{x}_j$, with 
$r_{ij,{\rm image}} = \vert \textbf{x}_{ij,{\rm image}} \vert$ (see section \ref{subsec:image vortices}). 
Furthermore, \textbf{$2 \pi \kappa$} is the quantum circulation, $\omega$ is the angular velocity of the inertial frame, and $V(\textbf{x}_i - \textbf{x}_k)$ is the pinning potential at $\mathbf{x}_i$ due to a pinning site located at $\mathbf{x}_k$, with $1 \leq k \leq N_{\rm pin} \neq N$ in general.
The first terms in equations \eqref{eq:x velocity} and \eqref{eq:y velocity} are the components of the fluid velocity at $\textbf{x}_i$ induced by the other vortices $1 \leq j \neq i \leq N$.
The second terms describe the motion due to image vortices, 
the third term is due to a rotating reference frame, the fourth term
describes the velocity induced by pinning sites, and the rotation matrix 
$\mathcal{R}_\phi$ in equation \eqref{eq:total velocity} describes the effect of dissipation.
We describe the origin and detailed form of each of these terms below.

\subsection{Vorticity-induced velocity}
\label{subsec:vortex_vortex}

The vorticity of a fluid is defined as the curl of the velocity field, i.e.,
\begin{equation}
\boldsymbol{\omega}(\mathbf{x}) = \nabla \times \mathbf{u}(\mathbf{x}) \, . 
\label{eq:vorticity equation}
\end{equation}
Given $\boldsymbol{\omega}(\mathbf{x})$, the velocity can be calculated using the Biot-Savart law,
\begin{equation}
\textbf{u}(\textbf{x}) = \frac{1}{4\pi} \int d^3\textbf{x}'
\frac{\boldsymbol{\omega}(\textbf{x}') \times (\textbf{x} - \textbf{x}')}
{\vert \textbf{x} - \textbf{x}' \vert^3} \, .
\label{eq:biot-savart}
\end{equation}
In this paper, we consider infinitely long, rigid vortices aligned parallel to each other. 
This describes a two-dimensional system exactly but ignores important three-dimensional phenomena, such as vortex tangles, vortex tension, and reconnection
\citep{Saffman1995}.

For a system of $N$ vortices moving only under vortex-induced motion in an unbounded fluid, equation \eqref{eq:biot-savart} reduces to
\begin{equation}
\frac{d \mathbf{x}_i}{dt} = \sum_{j\neq i}
\kappa \frac{\hat{\mathbf{z}}\times \mathbf{x}_{ij}}{r_{ij}^2} \, ,
\label{eq:2D biot-savart}
\end{equation}
for a vortex at point $\mathbf{x}_i$.
For a pair of vortices, equation \eqref{eq:2D biot-savart} describes counter-clockwise rotation of each vortex about the centroid of the pair.
The term $j = i$ is excluded in equation \eqref{eq:2D biot-savart} because there is no centre of rotation for an unbounded fluid: in the absence of other vortices, an initially stationary isolated vortex remains stationary regardless of its initial position.

Following 
\citet{Lin1941}, 
we note that \eqref{eq:2D biot-savart} is equivalent to Hamilton's equations of motion for the Hamiltonian
\begin{equation}
\mathcal{H} = \sum_{i=1}^N \sum_{j \neq i} \kappa \ln(r_{ij}) \, ,
\label{eq:hamiltonian}
\end{equation}
where the Cartesian coordinates $x_i$ and $y_i$ are the conjugate variables.

\subsection{Boundary conditions and image vortices}
\label{subsec:image vortices}

If the fluid is bounded by a container, the equations of motion
are modified in order to enforce the boundary condition that the normal component of the induced velocity at the boundary of the container vanishes.
This boundary condition can be solved using the method of images.
For a single vortex at polar coordinates $(r, \psi)$, in a cylindrical container 
of radius $R$ with $r < R$, the radial velocity at an arbitrary 
boundary point $(R, \theta)$ is
\begin{equation}
u_r = \frac{\kappa r \sin (\theta - \psi)}
{R^2 + r^2 - 2Rr \cos(\theta - \psi)} \, .
\label{eq:radial velocity}
\end{equation}
The radial component of the induced velocity at the boundary due to a vortex at polar coordinates $(R^2/r, \psi)$ is the same as equation \eqref{eq:radial velocity}, so the boundary condition $u_r = 0$ can be satisfied by placing an image vortex with opposite circulation 
(i.e. $\kappa \rightarrow -\kappa$) at $(R^2/r,\psi)$.  

With a circular boundary, the right-hand side of equation \eqref{eq:2D biot-savart} includes an extra term due to the image vortices,
\begin{equation}
-\sum_{j=1}^N \kappa 
\frac{\hat{\mathbf{z}} \times \mathbf{x}_{ij, {\rm image}}}
{r^2_{ij, {\rm image}}} \, ,
\label{eq:image velocity}
\end{equation}
where 
$\mathbf{x}_{ij,{\rm image}}$, and
$r_{ij, {\rm image}}$ are defined in section \ref{subsec:equations_of_motion}.
The sum over the image vortices does not exclude the $j = i$ term, because the presence of the boundary imposes a fixed centre, i.e. even a single vortex offset from the centre of the container rotates about the centre.

\subsection{Rotating frame}
\label{subsec:frame velocity}

We run simulations in a rotating frame, whose angular velocity $\omega$ is chosen initially such that the total circulation satisfies 
$\oint \mathbf{v}\cdot d\mathbf{l} = R^2 \omega = N \kappa$. 
The rotating frame enters through the third term in equations \eqref{eq:x velocity} and \eqref{eq:y velocity}.
The container and any pinning sites attached to the container (see section \ref{subsec:pinning velocity}) are stationary in the rotating frame.
Note that the container and the superfluid do not in general have the same angular velocity in the simulations.
The relation $R^2 \omega = N \kappa$ is chosen such that, in equilibrium (if reached hypothetically), the vortices and container co-rotate.

\subsection{Pinning}
\label{subsec:pinning velocity}

In a neutron star, nuclear lattice sites or magnetic flux tubes interrupt the smooth outward flow of vortices as the neutron star decelerates.
This effect is called `pinning'
\citep{Alpar1977}.
It arises due to quantum mechanical interactions between the vortex cores and the lattice sites or flux tubes
\citep{Link2009,Drummond2017,Drummond2018}.
In laboratory superfluids, pinning occurs at imperfections in the container
\citep{Tsakadze1980}.
In this paper, we consider a grid of pinning sites at positions $\mathbf{x}_k$ in the computational domain, which are stationary in the rotating frame described in section \ref{subsec:frame velocity}.
Pinning leads to a term $\sum_k V(\mathbf{x - x}_k)$ in equation \eqref{eq:hamiltonian}, where  $V(\mathbf{r}) = -V_0 f(r)$ is an attractive pinning potential, and $f(r)$ is some radially symmetric function,
e.g. a Gaussian. 
Evaluating the equations of motion for this extended Hamiltonian gives the fourth terms in equations \eqref{eq:x velocity} and \eqref{eq:y velocity}.
The terms describe clockwise rotation of vortices about the pinning sites, which counteracts the counter-clockwise rotation of vortex pairs about their centroids 
\citep{Acheson1990}.

\subsection{Dissipation}
\label{subsec:drag angle}

A vortex array has a tendency to minimize its free energy
[see e.g. \citet{Abrikosov1957, Fetter1965, Campbell1979}].
However, the motion described by equation \eqref{eq:hamiltonian}, is conservative; $\mathcal{H}$ is a constant of the motion.
Some dissipative mechanism is necessary to reach this lowest energy state.
\citet{Campbell1979} 
showed that the free energy minimum can be reached for vortices in an arbitrary initial configuration by moving the vortices along a vector parallel to the gradient of the free energy.
However, these authors were interested only in finding stable equilibrium states and did not suggest a physical mechanism for reaching these states.
We incorporate dissipation using the formalism developed for hydrodynamic descriptions of superfluid helium.
Phenomenological models of superfluids posit a dissipative interaction between the condensate and the viscous `normal' component that is analogous to drag [e.g.
\citet{Hall1956}]. 
\citet{Sedrakian1995}
showed that in this formalism the equations of motion for a vortex line in two dimensions can be written as a combination of induced motion due to the other vortices considered in aggregate, and a rotation through a `dissipation angle', whose value is related to the strength of the mutual friction between the inviscid and viscous components of the fluid.
In our model, we first evaluate the velocity of the vortices due to real vortices, image vortices, and pinning sites, then rotate the velocity vectors through a dissipation angle, $\phi$, related to thermodynamic properties of the superfluid, by multiplying with the rotation matrix $\mathcal{R}_\phi$ in \eqref{eq:total velocity}.
In section \ref{subsec:dissipation tests}, we verify that, in the absence of pinning, this procedure causes a vortex array initialized in an arbitrary configuration to eventually form an Abrikosov array which co-rotates with the frame of the simulation
\citep{Abrikosov1957,Campbell1979}.

\section{Numerical method}
\label{sec:methods}

We write a Python code that solves Equations \eqref{eq:total velocity}--\eqref{eq:y velocity} using a fourth-order explicit adaptive Runge-Kutta Cash-Karp (RKCK) method
\citep{Press1992}.
A vortex simulation begins by initializing the vortex locations, calculating the initial vortex velocities, then stepping the system forward in time.

\subsection{Feedback}
\label{sec:feedback}

In order to study avalanches, e.g. in neutron star applications, we consider a superfluid which is confined by a rotating container which decelerates. 
The dissipation mechanism discussed in section \ref{subsec:drag angle} causes the vortices to migrate outwards in order to maintain corotation.
As the vortices migrate outwards, the angular momentum of the superfluid decreases.
An equal and opposite angular impulse feeds back onto the container, 
slowing the spin-down rate.
The astrophysical details of the feedback mechanism lie outside the scope of this paper.
We assume here for simplicity that it is instantaneous and lossless
\citep{Fulgenzi2017,Carlin2020}.

Let $L_c = I_c \Omega_c$ and $L_s = I_s \Omega_s$ be the angular momentum of the container and the superfluid respectively and $I_c$ and $I_s$ be their respective moments of inertia.
The evolution equation for $\Omega_c$ is
\begin{equation}
\frac{d \Omega_c}{dt} = N_{\rm ext} - I_{\rm rel}\frac{d \Omega_s}{dt} \, ,
\label{eq:angular momentum evolution}
\end{equation}
with $I_{\rm rel} = I_s/I_c$, where $N_{\rm ext}$ is the external spin-down torque divided by $I_c$.
For a rotating superfluid, the angular momentum within radius $R$ is 
\citep{Fetter1965}
\begin{equation}
L_s = k \sum_{i=1}^N (R^2 - r_i^2) \, ,
\label{eq:angular momentum symmetric}
\end{equation}
where $k$ is a constant with units of kg s$^{-1}$, and $r_i$ is the radial coordinate of the $i$-th vortex.
If we assume axisymmetry, equation \eqref{eq:angular momentum symmetric} reduces to $L_s = kN(R^2 - \langle r^2 \rangle)$, where $\langle r^2 \rangle$ denotes the average over the vortices of the square of the radial coordinate.	
For a uniform vortex array we obtain $\langle r^2 \rangle = R^2/2$,  $L_s = kNR^2/2$.

The spin-down and feedback procedure works as follows. 
At each time step:
\begin{enumerate}[label=(\arabic*)]
\item compute $\Omega_s$ from equation \eqref{eq:angular momentum symmetric};
\item increment $\Omega_c$ by $N_{\rm ext} \Delta t$;
\item update the vortex positions at the new time step according to equation \eqref{eq:total velocity};
\item compute $\Omega_s$ from equation \eqref{eq:angular momentum symmetric};
\item decrement $\Omega_c$ by $I_{\rm rel} \Delta \Omega_s$.
\end{enumerate}
Updating $\Omega_c$ only at the beginning and end of the time step does not strictly adhere to the fifth-order RKCK scheme used to solve equations \eqref{eq:x velocity} and \eqref{eq:y velocity}.
However, it is a good approximation if we assume that the time-scale over which vortices adjust to changes in the angular velocity of the container is much faster than other relevant time-scales.

\subsection{Dimensionless coordinates}
\label{subsec:coordinates}
We run our simulations in a dimensionless coordinate system where $\kappa=1$ and the fundamental length unit is one.
All other quantities, such as the time and velocity, are defined through these quantities.
Our system of equations does not include any mass terms, as the vortex `particles' are features in the velocity field and intrinsically massless. 
Other quantities involving mass, such as angular momenta and moments of inertia, only enter our equations of motion in dimensionless units such as $I_{\rm rel}$.

\subsection{Glitch finding}
\label{subsec:glitch finding}

The main observables in pulsar glitch studies are the glitch sizes and inter-glitch waiting times.
Pulsar radio emission is thought to be phase-locked to the rigid stellar crust, so both observables relate directly to the evolution of $\Omega_c(t)$.
The procedure we use in order to extract sizes and waiting times from our simulations is as follows.
The angular velocity of the container, $\Omega_c$, is recorded at each time step.
When the sign of $d\Omega_c / dt$, the rate of change in $\Omega_c$ between successive time steps, is positive, we flag the epoch as $t_{\rm g}$.
We then compute the cumulative increase in $\Omega_c$ between $t_{\rm g}$ and the next epoch when $d\Omega_c / dt$ becomes negative, $t_{\rm end}$.
The cumulative increase in $\Omega_c$ between $t_{\rm g}$ and $t_{\rm end}$ is recorded as the size of the glitch at epoch $t_{\rm g}$.
The waiting times are computed as the difference between successive epochs $t_{\rm g}$.
We do not compute a waiting time for the first glitch, because the simulation takes a while to build up enough stress to trigger the first vortex avalanche, i.e., it is not yet stationary statistically.

We note in passing that the above method of glitch finding broadly mirrors how astronomical glitches are detected.
Most pulsars are monitored sporadically, with typical cadences of days to months. 
Hence pulsar glitches are almost never detected in real time [though see 
\cite{Palfreyman2018}].
Instead, they are identified through secular changes in timing residuals relative to a glitch-free timing model, which incorporates the astrometric and rotational evolution of the pulsar between observations
\citep{Lorimer2004}.
 
\section{Validation}
\label{sec:tests}

We verify that the solver described in section \ref{sec:methods} works as intended by reproducing two classical results in point vortex dynamics
\citep{Acheson1990}: the stability of a vortex ring (section \ref{subsec:ring tests}), and the formation of an Abrikosov array (section \ref{subsec:dissipation tests}).

\subsection{Stability of a vortex ring}
\label{subsec:ring tests}

The study of vortex motion in two dimensions goes back to the work of Helmholtz and others in the 19th Century.
An important early result, first discussed by
\citet{Thomson1883}
and later proved by 
\citet{Havelock1931}, 
describes the motion of $N$ point vortices evenly spaced around a ring of radius $R$ in an infinite medium.
In the absence of dissipation, the vortices rotate uniformly about the centre of the ring, with angular velocity $\omega_N = \kappa(N-1)/ (2 R^2)$.
For $N < 7$, the system is stable to small perturbations and maintains circular motion indefinitely. 
For $N = 7$, the system is neutrally stable.
For $N > 7$, the ring configuration is unstable, and a transition from circular to chaotic motion is unavoidable.
\citet{Havelock1931} performed a full linear stability analysis of this problem, including an extension to systems involving inner and outer boundaries, as well as counter-propagating vortex rings.

In order to test the solver, we reproduce three results: 
(i) the angular velocity of each vortex in the ring is proportional to the number of vortices, with 
$\omega_N = \kappa(N-1)/(2 R^2)$; 
(ii) rings with $N < 7$ are stable indefinitely while those with $N > 7$ result in chaotic motion; and
(iii) the $N = 7$ case is metastable, transitioning from stable to chaotic motion when perturbed.
We perform a suite of simulations where we solve equation \eqref{eq:2D biot-savart} for $2\leq N \leq 10$, and $0 \leq t \leq 200 \pi / \omega_N$ (i.e. 100 rotation periods). 
\begin{figure}
\includegraphics[scale=0.5]{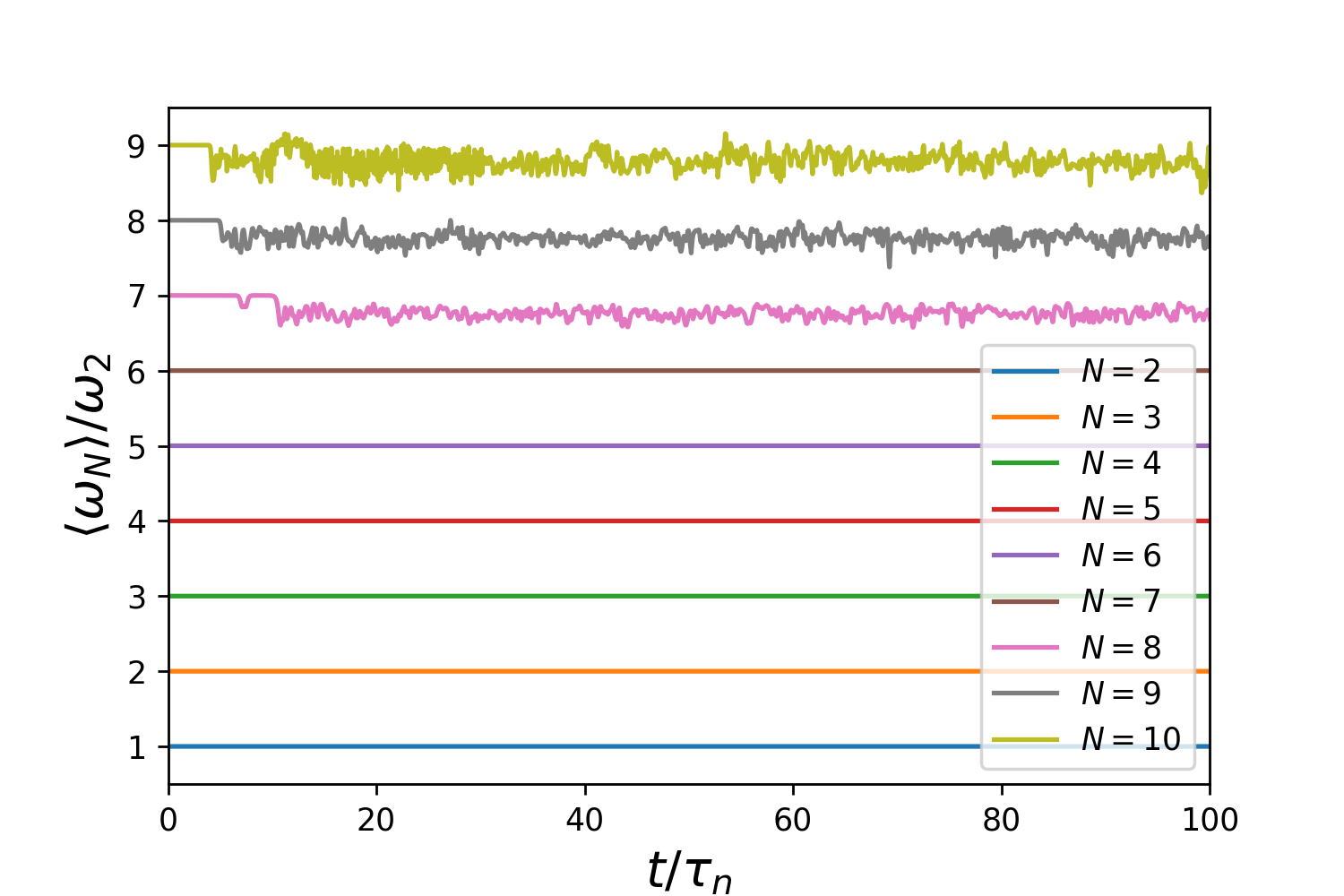} \\
\includegraphics[scale=0.5]{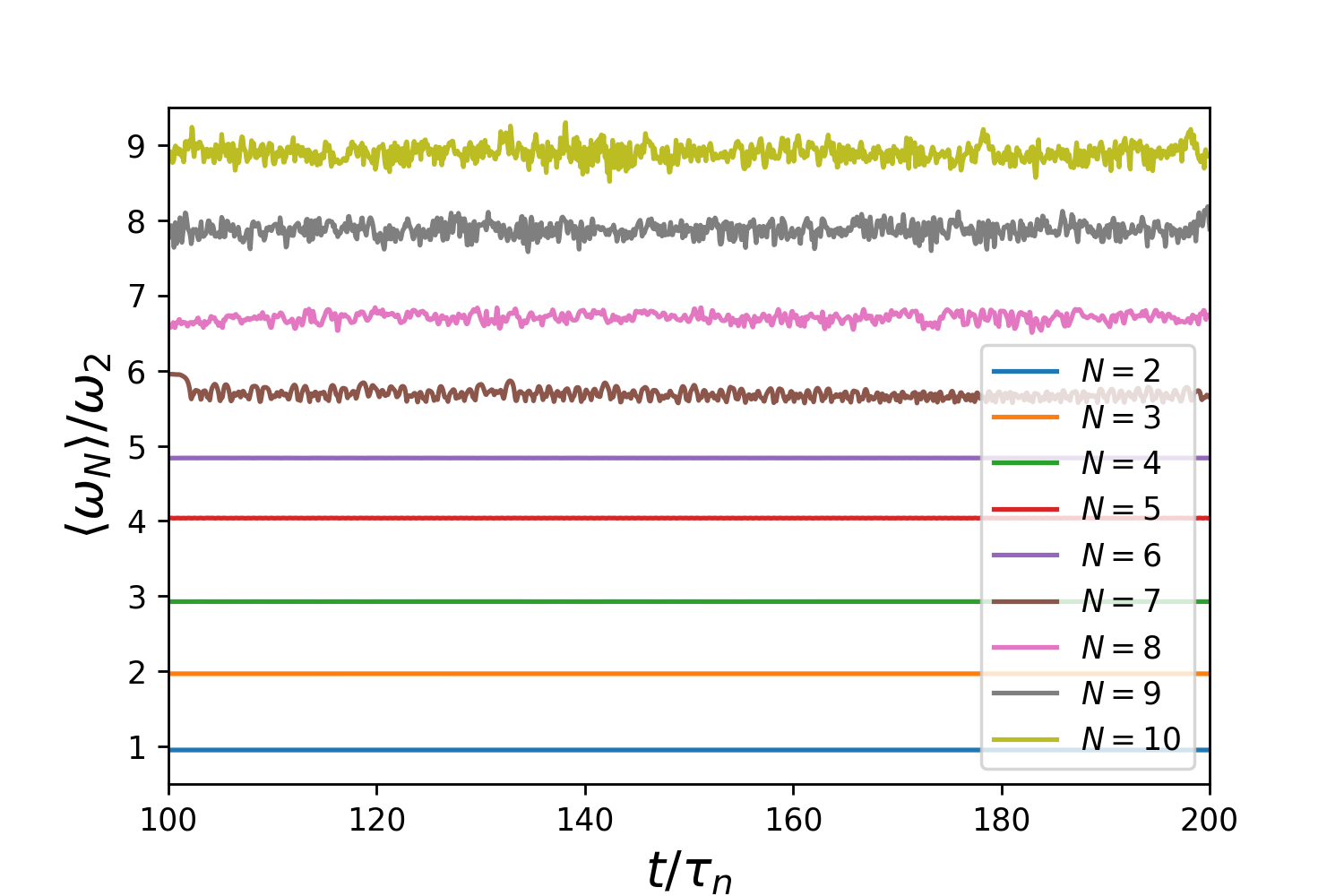}
\caption{Vortex-averaged angular velocity of $N$ vortices, normalised by $\omega_2$, versus time, normalised by $\tau_N$, for $2 \leq N \leq 10$.
Top panel: vortex positions are initialised at the vertices of a regular polygon.
Bottom panel: vortices are given a random radial displacement of $\pm 5\times 10^{-2}R$ relative to the centroid of the initial polygon.
Note: the time axis starts at $t/\tau_N=0$ (100) in the top (bottom) panels.
}
\label{fig:ringtest}
\end{figure}
In Figure \ref{fig:ringtest}, we show the angular velocity of the vortices, averaged over all vortices, normalised by $\omega_2$, versus time, in units of the rotation period of the vortex ring $\tau_N = 2 \pi / \omega_N$.
In the top panel, the vortices are placed initially at the vertices of a regular polygon and are evolved for $0 \leq t/\tau_N \leq 100$.
In the bottom panel, we perturb the system by displacing each vortex radially by $\pm 5 \times 10^{-2} R$ (with $+$ or $-$ chosen at random for each vortex) relative to the centroid of the initial polygon and evolve the system for $100 \leq t/\tau_N \leq 200$.
Figure \ref{fig:ringtest} shows that $\langle \omega_N \rangle$ increases linearly with $N$, as expected.
In the top panel, for $N \leq 7$, $\langle \omega_N \rangle / \omega_2$ is constant with time.
For $N > 7$, the motion becomes disordered at $t \approx 10 \tau_N$ and $\langle \omega_N \rangle / \omega_2$ fluctuates noisily.
In the bottom panel, after perturbing the vortex positions initially, the rings with $N < 7$ and $N > 7$ are unaffected, but the metastable $N=7$ ring comes to resemble the $N>7$ cases.

\begin{figure}
\begin{center}
\includegraphics[scale=0.5]{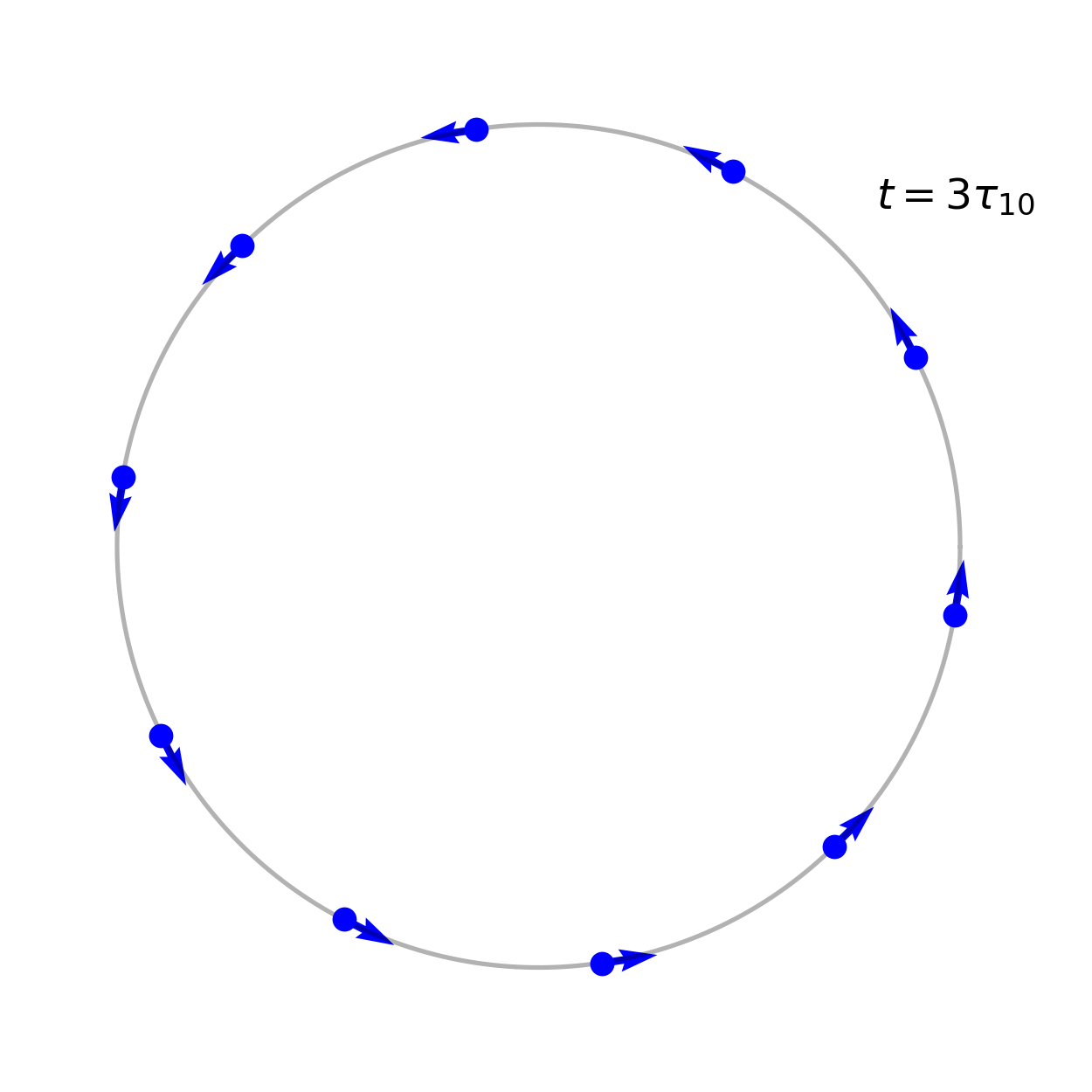} \\
\includegraphics[scale=0.5]{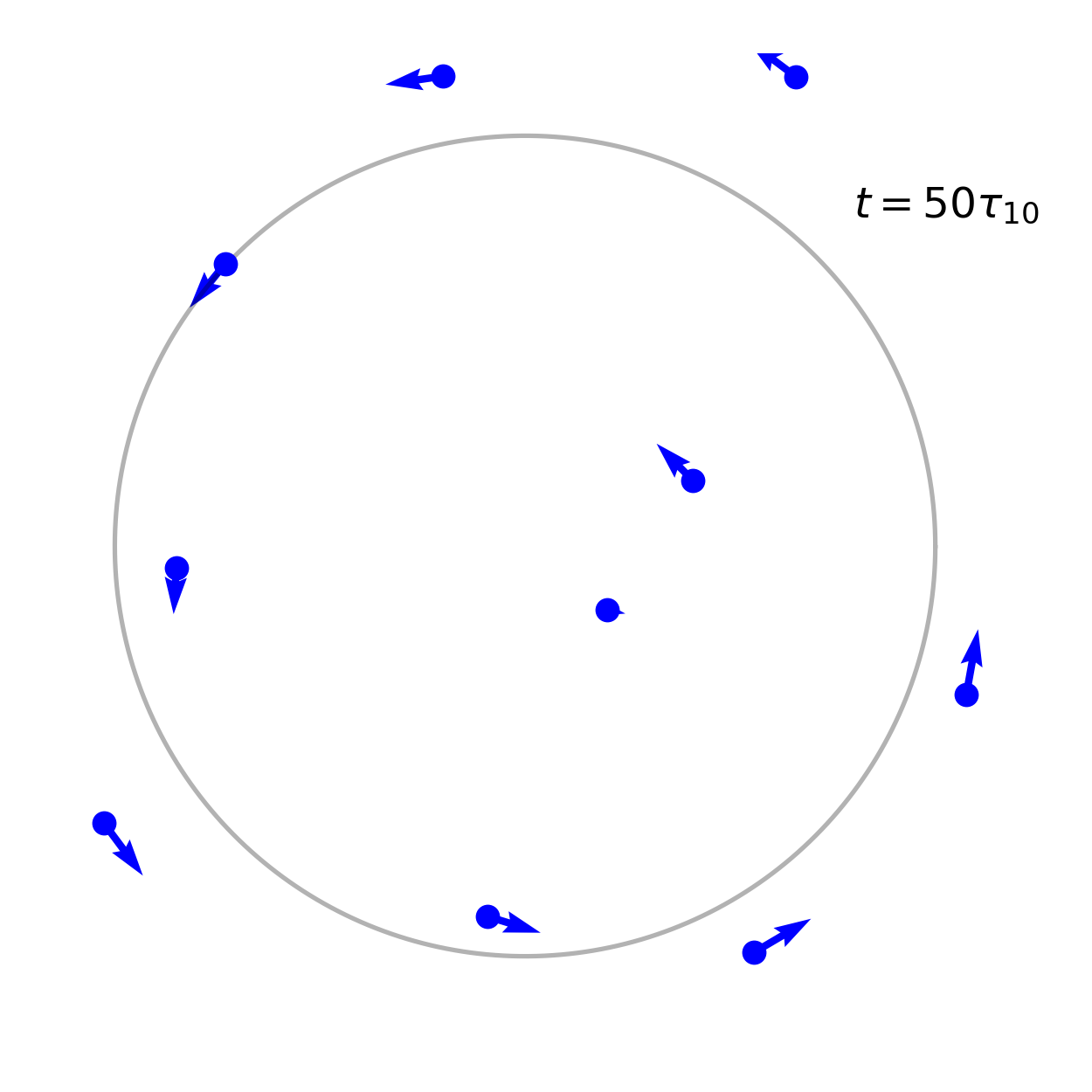}
\caption{Vortex positions (blue circles) and velocities (indicated by blue arrows) of the $N=10$ ring initialised in a regular polygon at $t = 3\tau_{10}$ (top panel) and at $t=50\tau_{10}$ before any radial perturbation at $t=100\tau_{10}$ (bottom panel).}
\label{fig:order_vs_chaos}
\end{center}
\end{figure}

To further illustrate the transition from ordered to disordered motion, in Figure \ref{fig:order_vs_chaos} we show the state of the 10-vortex ring at $t=3\tau_{10}$ and $t=50 \tau_{10}$.
The top panel shows the ordered behaviour at the beginning of the simulation.
The vortices are all equidistant from the centre of the circle, and their velocities are equal and tangent to the circle.
In the bottom panel, the positions and velocities are randomized.

\subsection{Dissipation and Abrikosov array formation}
\label{subsec:dissipation tests}

\citet{Abrikosov1957}
showed that flux tubes in a type II superconductor arrange themselves in a triangular array, as long as the system is unbounded.
The same is true of flux tubes in a rotating superfluid such as helium II
\citep{Campbell1979}. 
When the vortices are confined to a cylindrical container, e.g. superfluid helium in a rotating bucket, the array is not exactly triangular near the boundary
\citep{Campbell1979}.

In order to verify that the dissipation mechanism in section \ref{subsec:drag angle} is working correctly, we show (i) that 
$\mathcal{H}$ converges for any initial vortex configuration and value of dissipation angle $\phi$, and (ii) that the final value of $\mathcal{H}$ corresponds to an approximately triangular array which co-rotates with the container.
We perform a suite of simulations solving equations \eqref{eq:x velocity} and \eqref{eq:y velocity} for an array of 100 vortices, choosing $\phi \in \langle 0.01,0.02,0.05,0.1,0.2,0.5 \rangle$, to study convergence as a function of dissipation angle. 
The initial vortex positions are either drawn at random from a uniform spatial distribution or spaced uniformly around a ring of radius $R$ (cf section \ref{subsec:ring tests}).
We also perform 10 simulations with $\phi=0.1$, with different initial vortex positions drawn at random from a uniform spatial distribution to study the ensemble statistics.
We run each simulation for $t \approx 10^4 T_0$, where $T_0$ is the rotation period of the container, determined by
the Feynman condition $T_0 = 2 \pi R^2 / N \kappa$.
We run the simulations in a rotating frame with $\omega = 2 \pi / T_0$.
While the terms involving $\omega$ in equations \eqref{eq:x velocity} and \eqref{eq:y velocity} confine vortices within $r<R$ at equilibrium, the vortices often overspill the boundary transiently at early times.
When image vortices are present, the overspilt vortices annihilate, and leave the simulation.
Because the initial positions are randomized, the number of vortices that survive is different in each simulation, and $\mathcal{H}$ does not reliably converge to the same value each time.
In the simulations described in this section, where we are mostly interested in the array configuration at equilibrium, we do not enforce the boundary condition in order to keep the vortex number constant.
All the simulations described in section \ref{sec:results} below include image vortices.
\begin{figure}
\begin{center}
\includegraphics[scale=0.4]{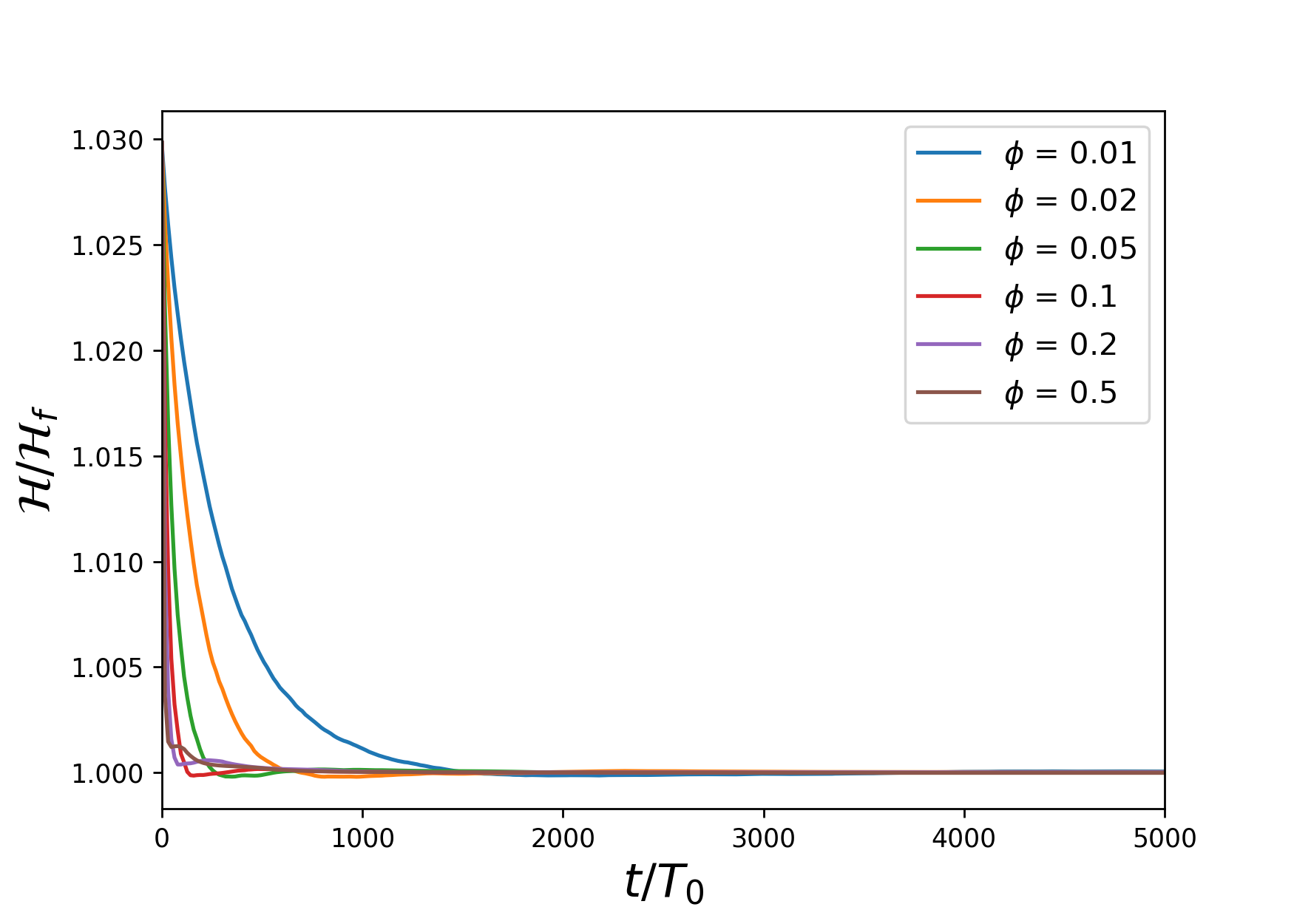} \\
\includegraphics[scale=0.4]{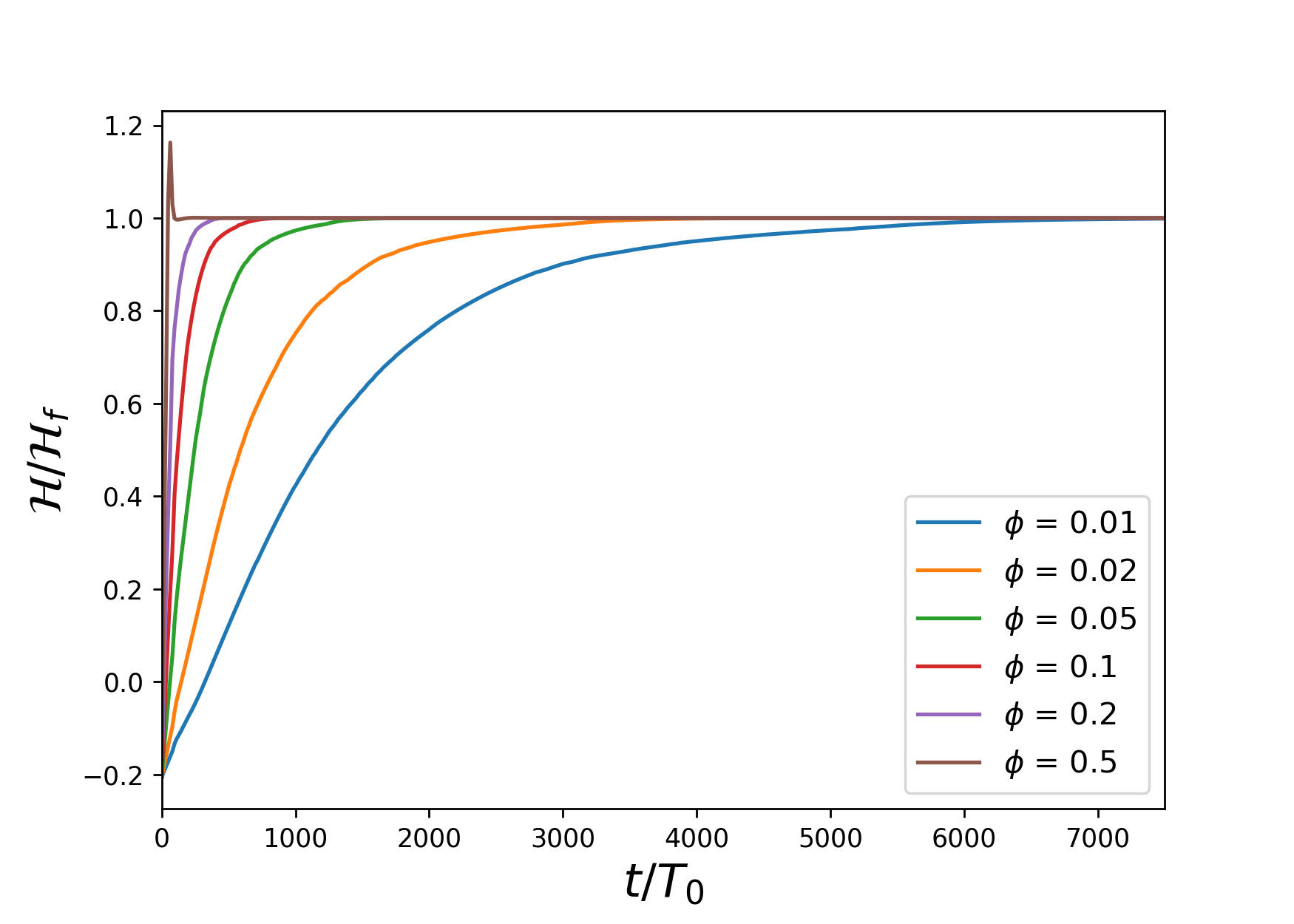} \\
\includegraphics[scale=0.4]{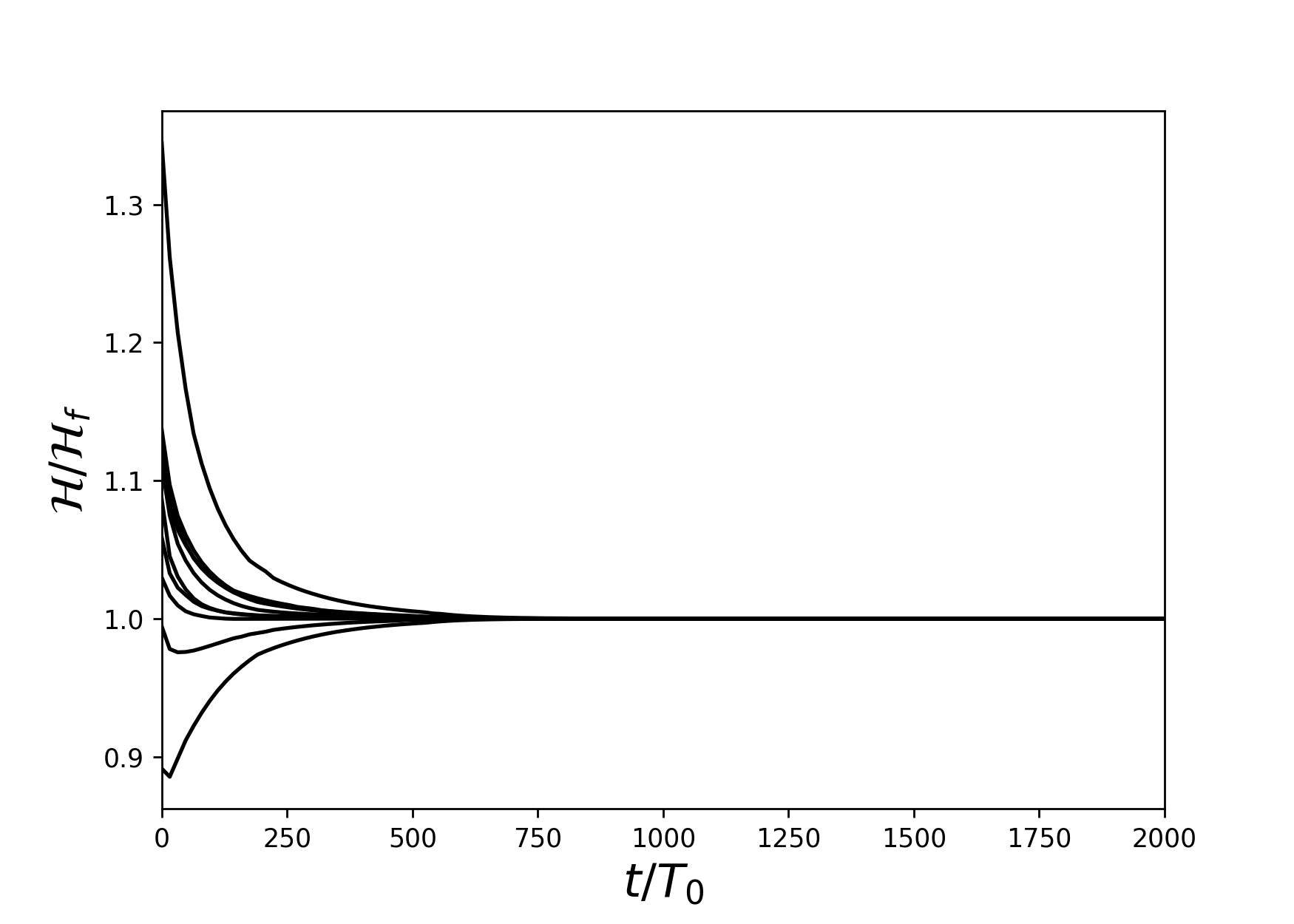} \\ 
\includegraphics[scale=0.4]{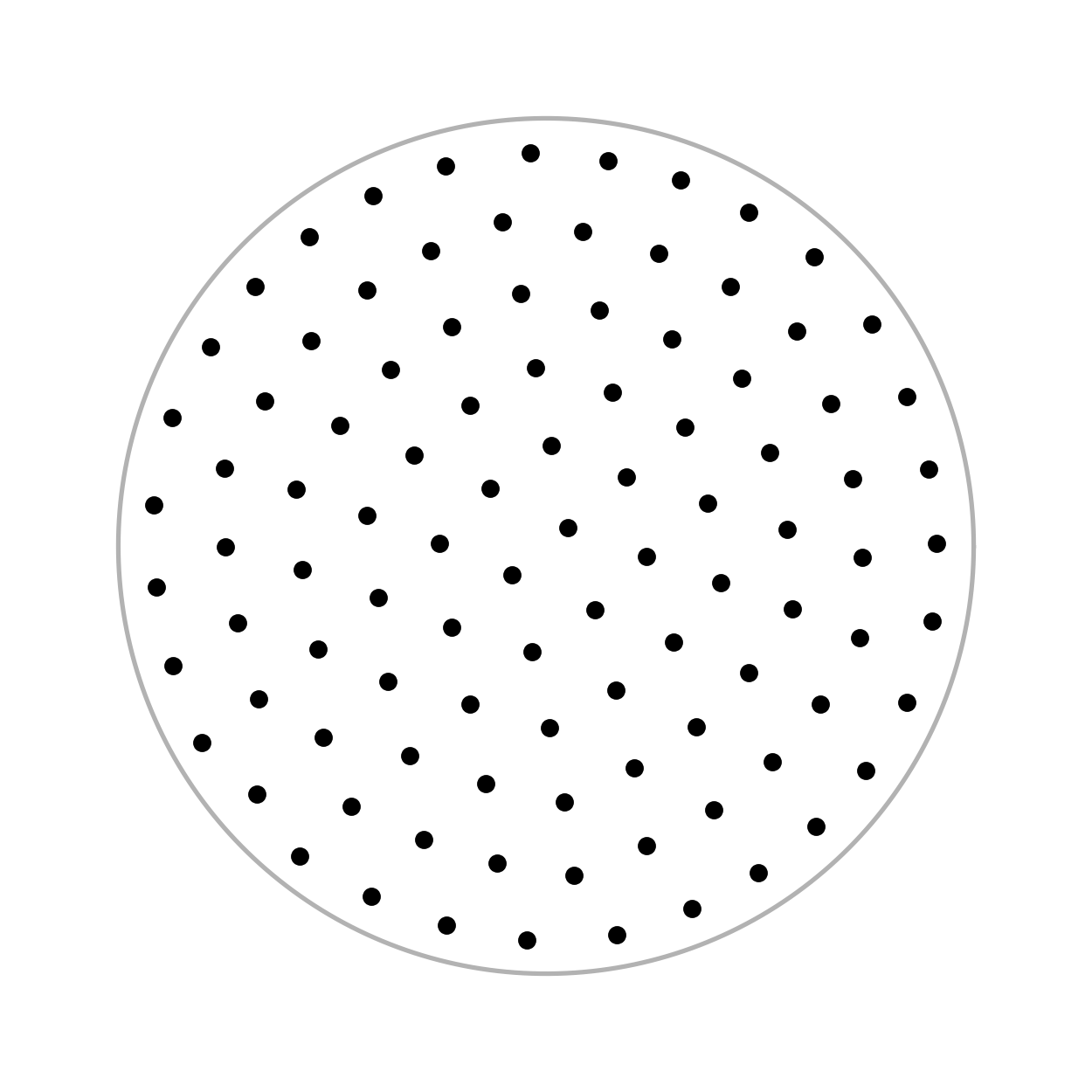}
\end{center}
\caption{Convergence of the vortex array from an arbitrary initial configuration to a triangular array.
Top panel: Hamiltonian $\mathcal{H}$ [equation \eqref{eq:hamiltonian}] versus time for 100 vortices initialised with random positions drawn from a uniform spatial distribution and evolved according to equation \eqref{eq:total velocity} 
(ignoring pinning and image vortices) for $0 \leq t \lesssim 10^4 T_0$, where $T_0$ is the rotation period of the container. 
Each simulation is initialised in the same configuration, with a different value of the drag angle 
$\phi \in \langle 0.01,0.02,0.05,0.1,0.2,0.5 \rangle$ (see legend).
Second panel: as for the top panel, but with the vortices initially situated equidistantly around a ring of radius $R$.
Third panel: $\mathcal{H}$ versus $t$ for 10 random uniform-density initial configurations with $\phi = 0.1$.
Bottom panel: Vortex positions at the end of one of the simulations in the third panel. 
}
\label{fig:convergence plots}
\end{figure}

Figure \ref{fig:convergence plots} graphs $\mathcal{H}/\mathcal{H}_{\rm f}$ versus $t$, where $\mathcal{H}_{\rm f}$ is the value of $\mathcal{H}$ at the end of the simulation with $\phi = 0.5$ and random initial vortex positions. 
The top two panels show $\mathcal{H}$ versus $t$ for $0.01 \leq \phi \leq 0.5$.
In the top panel the initial vortex positions are drawn at random from a uniform spatial distribution.
In the second panel the initial vortex positions are the vertices of a regular polygon.
The third panel shows $\mathcal{H}$ versus $t$ for 10 random uniform-density initial configurations with $\phi = 0.1$.
The bottom panel shows the vortex configuration at the end of one of the simulations shown in the third panel.

Figure \ref{fig:convergence plots} verifies properties (i) and (ii) discussed in the first paragraph of section \ref{subsec:dissipation tests}.
The bottom panel shows that the vortices settle to an approximately triangular array, which is stationary in the corotating frame.
The top three panels verify convergence, with $\mathcal{H} \rightarrow \mathcal{H}_{\rm f}$ in all simulations.
In the top two panels, we see that convergence takes longer for lower values of $\phi$, with 
$\vert \mathcal{H}/\mathcal{H}_f - 1 \vert > 10^{-3}$ for $t < 1.5 \times 10^{3} T$ ($\phi = 0.01$, top panel), and
$\vert \mathcal{H}/\mathcal{H}_f - 1 \vert > 10^{-3}$ for $t < 130 T$ ($\phi=0.5$, second panel).
Across all simulations the final value of $\mathcal{H}$ differs from $H_f$ by $< 0.1 \%$.
In the second panel, and in some of the simulations shown in the third panel, we see $\mathcal{H} < \mathcal{H}_f$ at early time.
This happens when some of the vortices enter the region $r>R$ temporarily, as discussed above.

\section{An Astrophysical Example: Neutron Star Glitches}
\label{sec:results}
In this section, we report the results of a suite of simulations done with the solver described in section \ref{sec:methods}.
The simulations investigate the far-from-equilibrium phenomenon of vortex avalanches driven by deceleration of the vessel containing the superfluid.
The results are compared with 
analogous numerical experiments involving Gross-Pitaevskii simulations, which reveal the microphysical knock-on mechanisms mediating the spatially correlated dynamics in vortex avalanches 
\citep{Warszawski2011,Warszawski2012,Warszawski2013,Melatos2015}.
The simulations form part of an idealized model of neutron star glitches triggered by vortex avalanches in a neutron superfluid coupled to a rigid stellar crust 
\citep{Anderson1975,Melatos2008,Haskell2015,Howitt2016}.

\subsection{Set up}
\label{subsec:simulation_description}
We initialise an ensemble of point vortices within a circular container of radius $R$ containing a square array of Gaussian pinning potentials, with the initial vortex positions drawn at random from a uniform spatial distribution.
We evolve the ensemble according to equation \eqref{eq:total velocity} with spin down and feedback turned off, until all of the vortices are pinned.
We then turn on feedback and spin down and resume evolving the system according to equations \eqref{eq:total velocity} and \eqref{eq:angular momentum evolution}.
The various input parameters and the meaning of each are explained in Table \ref{tab:parameters}.
\begin{table}
	\begin{tabular}{|c|c|}
	\hline 
	Parameter  & Physical meaning \\ 
	\hline
	$N_v$ & Number of vortices \\
	$R$ & Radius of container \\
	$\Delta t$ & Time step \\
	$I_{\rm rel}$ & Ratio of superfluid/crust moments of inertia \\
	$V_0$ & Pinning strength \\
	$a$ & Pinning site separation \\
	$\xi$ & Characteristic width of pinning sites \\ 
	$\phi$ & Dissipation strength \\
	$N_{\rm ext}$  & Spin-down rate \\	
	\hline 
	\end{tabular}
\caption{Summary of input parameters for the vortex avalanche simulations. 
Values used in the simulation are discussed in the text in sections \ref{subsec:simulation_description} and \ref{subsec:variations}.
} 
\label{tab:parameters}
\end{table}
The default parameters we use are $N_v=2000$, $R = 10$, $\Delta t = 0.1 T_0$ (where $T_0 = 2 \pi R^2 / N \kappa$ is the initial rotation period of the container), $I_{\rm rel} = 1$, $V_0 = 2000$, $a = 0.01 R$ (corresponding to a ratio of pinning sites/vortices $\approx 10$), $\xi=0.001 R$, $\phi=0.1$ radians, and $N_{\rm ext} = -5 \times 10^{-4} \Omega_0 / T_0$.
We run for $2 \times 10^5$ time steps, so that, in the absence of feedback, the container's period doubles to $2T_0$.
In order to obtain a statistically useful number of glitches for analysis, we perform three simulations with the same parameters, each beginning from a different random initial vortex configuration, and aggregate the glitches.
We have also performed a smaller suite of simulations with $N_v=5000$ as a cross-check. 
These produced qualitatively similar results to the $N_v=2000$ simulations described here, however, due to computational resource and time constraints we are unable to run them long enough to produce the number of glitches necessary for meaningful statistical analysis.

\subsection{Avalanche dynamics}
\label{subsec:avalanches}

We test for avalanche behaviour by examining an ensemble of identically initialised simulations for evidence of collective vortex motion mediated by one or more knock-on mechanisms
\citep{Warszawski2013}.
One such form of evidence is obtained by examining a movie of the vortex motion visually.
Is the vortex unpinning triggered at one point and does it then spread in a connected way (avalanche), or does it occur simultaneously at multiple locations (not an avalanche)?
Another form of evidence is obtained from the event statistics.
Are the sizes and waiting times distributed according to a power law and exponential respectively, as in avalanche-dominated self-organised critical systems
\citep{Jensen1998,Melatos2008}?
Owing to computational cost, we analyse an ensemble of three simulations in this section; a fuller study will be conducted elsewhere.
Identically initialised means that the macroscopic system variables ($N_v, R, \Delta t, I_{\rm rel}, a, \xi, \phi, N_{\rm ext}$) are the same in all three runs, but the initial vortex positions are selected randomly from a spatially uniform distribution.

\begin{figure}
\includegraphics[scale=0.6]{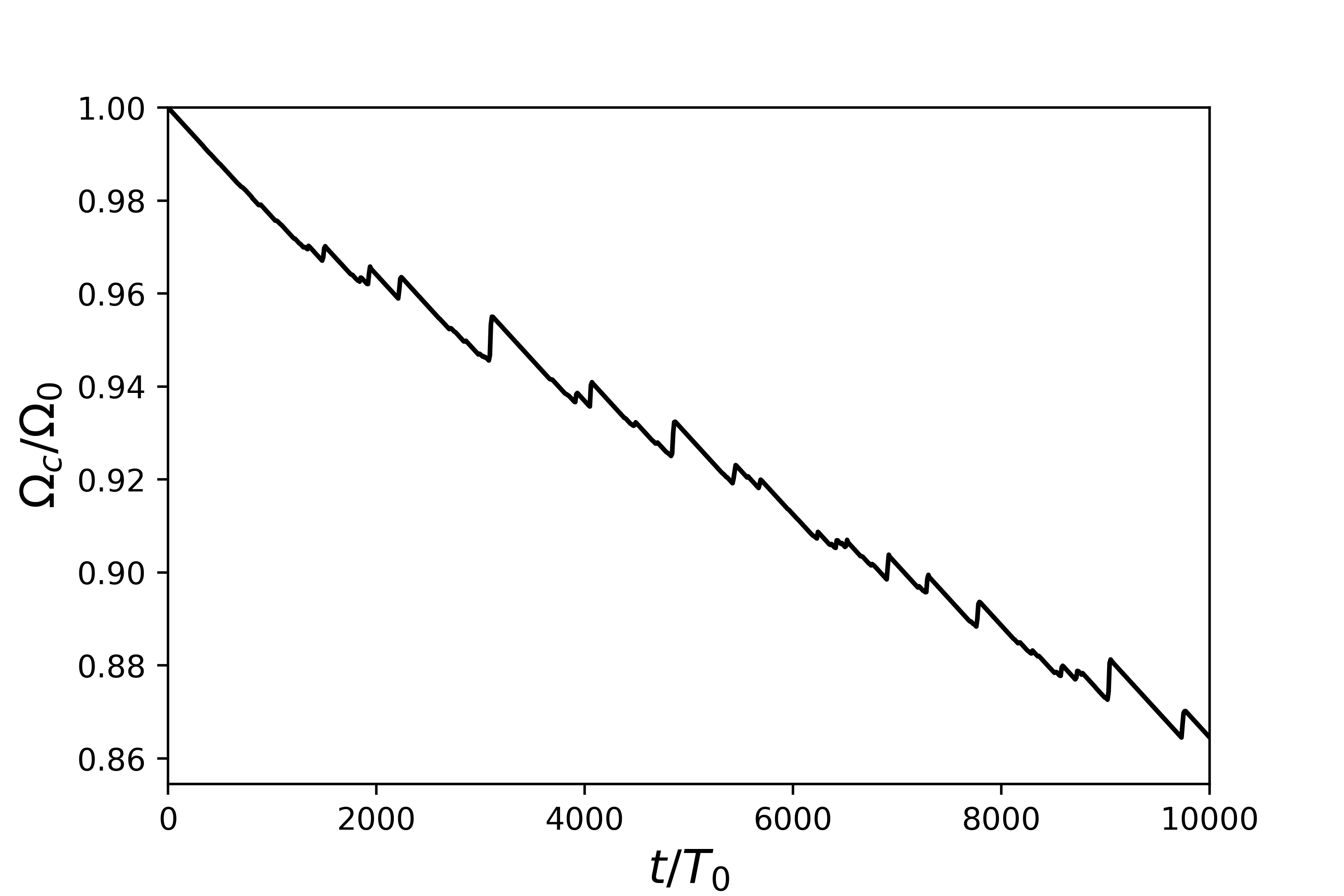}
\caption{Container angular velocity $\Omega_C$, normalized by its initial value, as a function of time over $10^4$ rotation periods. 
This simulation uses the default parameters listed in section \ref{subsec:simulation_description}.}
\label{fig:spindown_plot}
\end{figure}

Figure \ref{fig:spindown_plot} shows the evolution of the container's angular velocity, $\Omega_C(t)$, for $0 \leq t/T_0 \leq 10^4$ for one of the simulations with the default parameters.
Initially, the spin down is smooth and monotonic; the vortices remain pinned, as stress builds up. 
At $t \approx 10^3 T_0$, the first of several small spin-up events (glitches) occurs, corresponding to collective unpinning of vortices.
These continue to occur spasmodically throughout the remainder of the simulation, with 99 events detected by the glitch-finding algorithm (40 events in the time span plotted in Figure \ref{fig:spindown_plot}), ranging in size $\Delta \Omega_C$ over 
$7.1 \times 10^{-7} \leq \Delta \Omega_C/\Omega_0 \leq 4.5 \times 10^{-3}$.
For 
$0 \leq t/T_0 10^3$,
when the superfluid is effectively decoupled from the container, the spin-down rate (determined by a least-squares fit) is $-4.8 \times 10^{-4} \Omega_0 / T_0$, and reduces to $-2.5 \times 10^{-4} \Omega_0 / T_0$ during $10^3 \lesssim t/T_0 \lesssim 2 \times 10^4$.
Once the glitches begin, they maintain the vortices near the critical unpinning threshold, repeatedly albeit spasmodically coupling the superfluid to the container, effectively doubling the inertia of the system ($I_{\rm rel} = 1$).

\begin{figure}
\begin{center}
\includegraphics[scale=0.19]{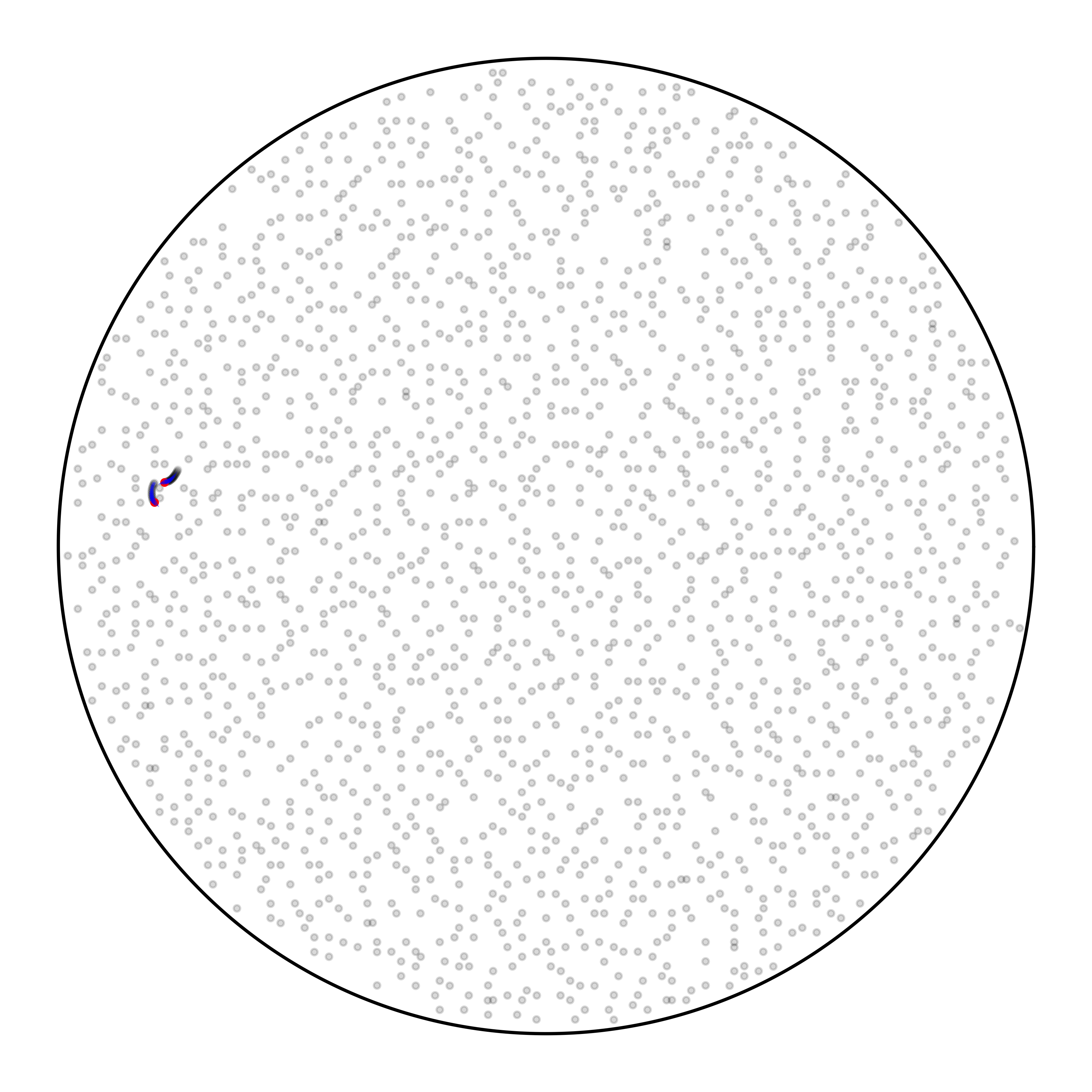}\\
\includegraphics[scale=0.19]{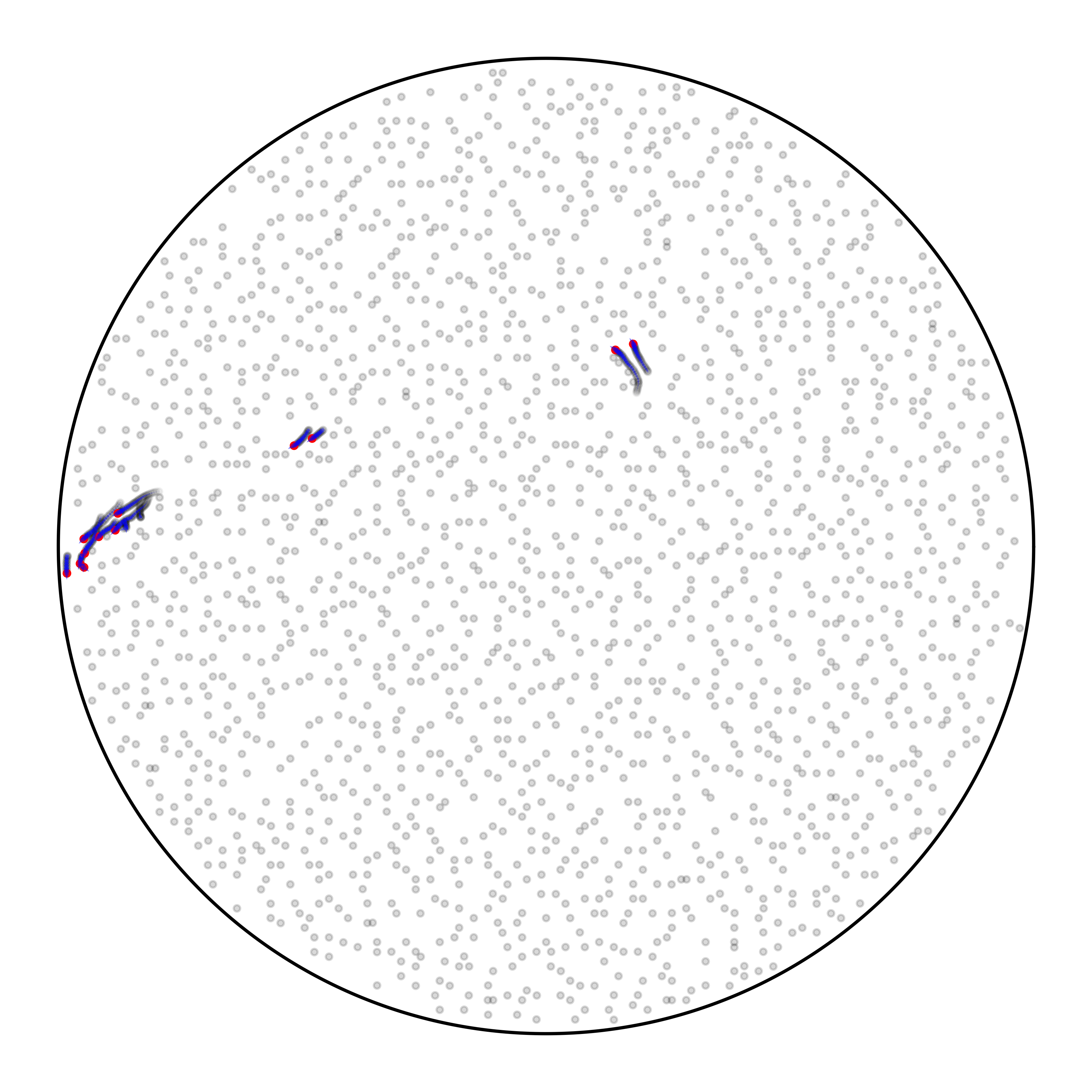}\\
\includegraphics[scale=0.19]{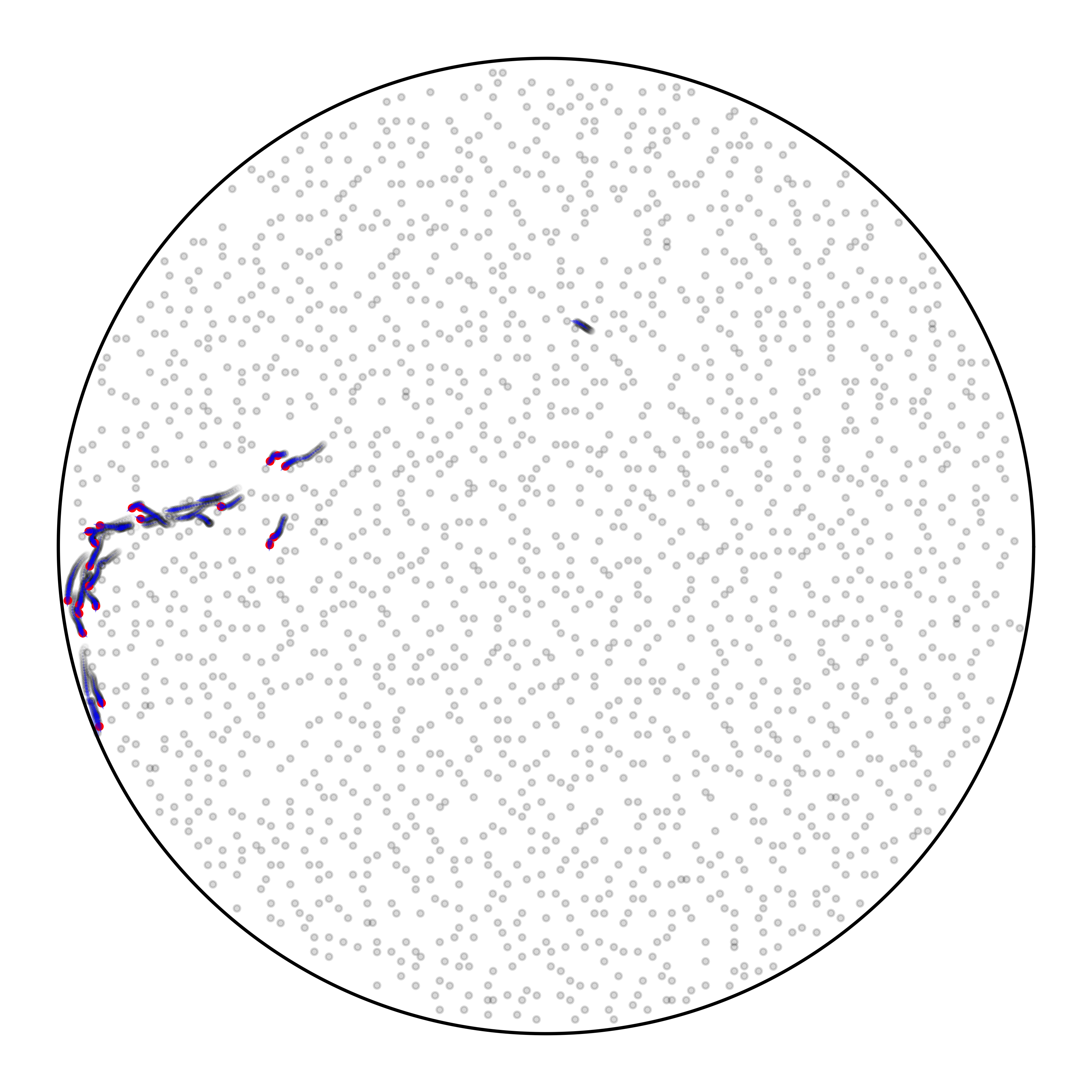}\\
\includegraphics[scale=0.19]{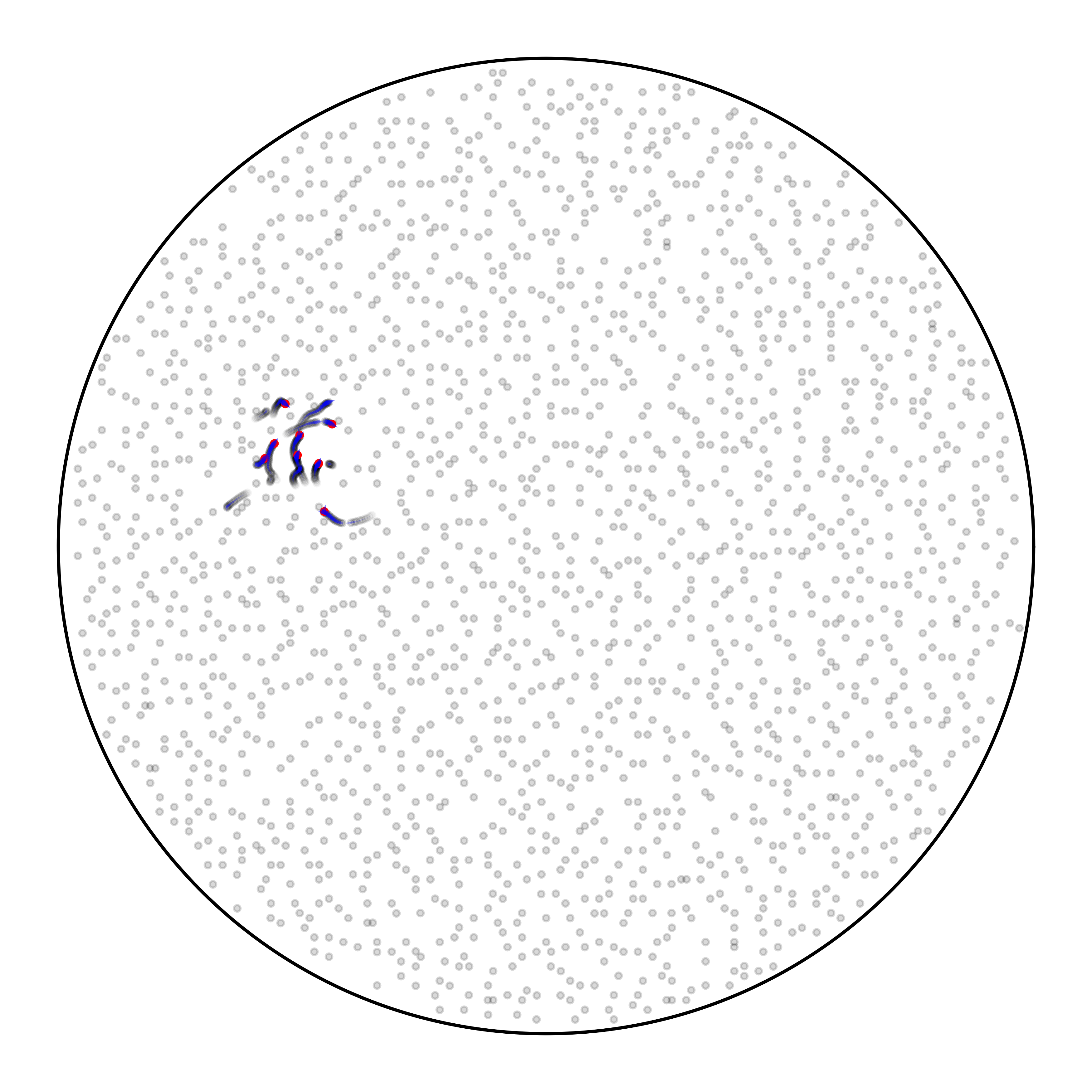}\\
\includegraphics[scale=0.19]{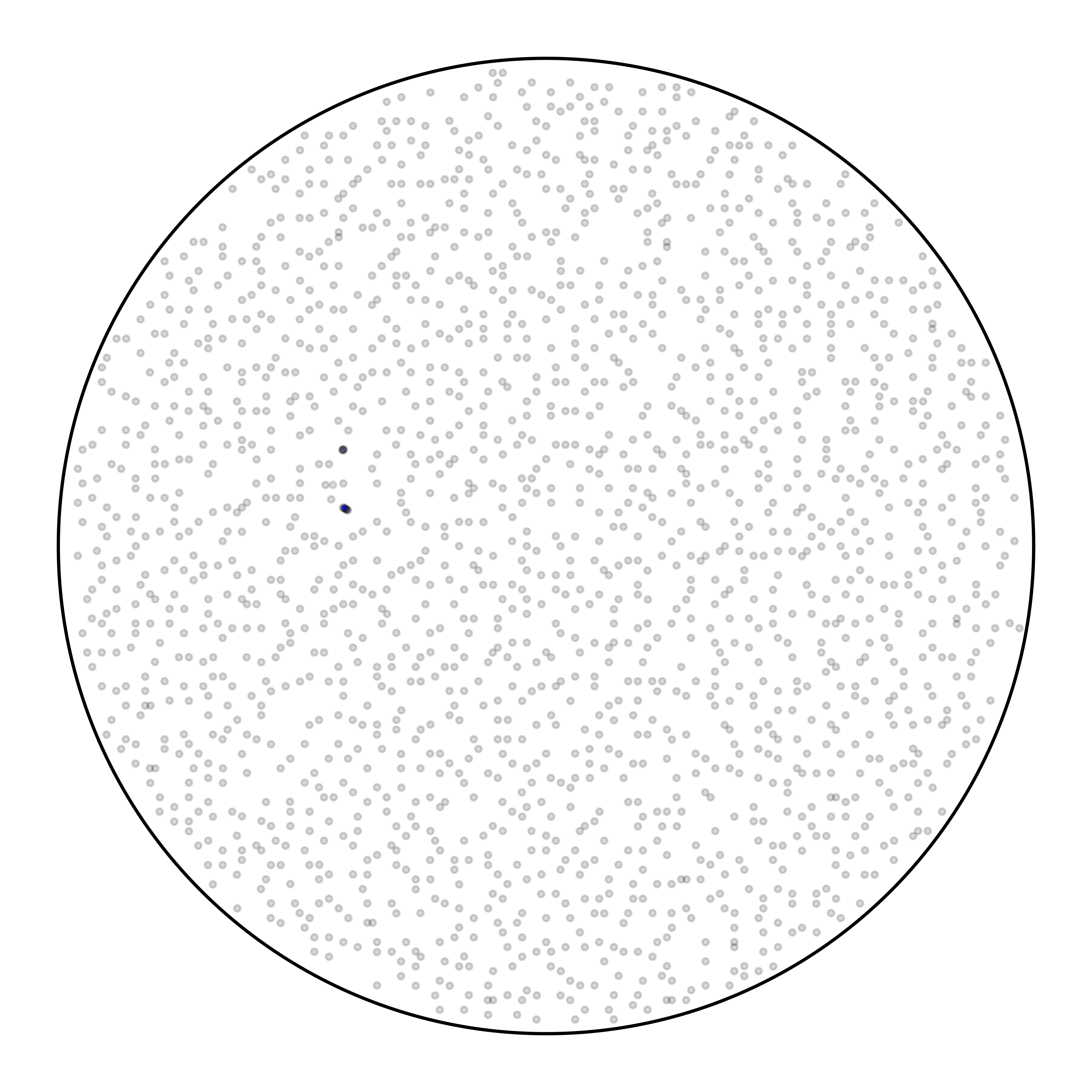}\\
\includegraphics[scale=0.4]{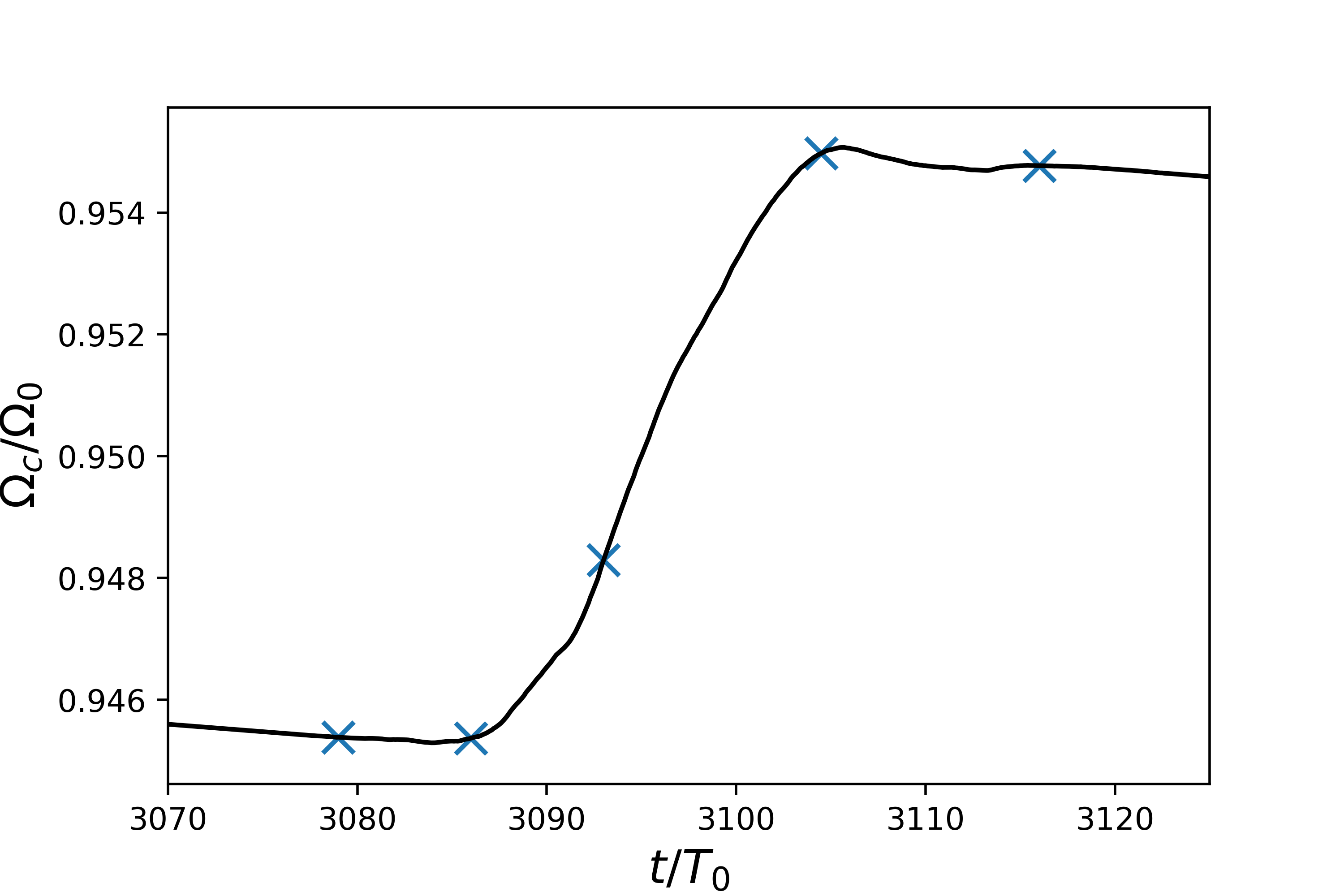}
\end{center}
\caption{Vortex motion during an avalanche. 
Top five panels are snapshots at (top to bottom) $t/T_0 =$ 3079, 3086, 3093, 3104.5, 3116. 
Grey dots show stationary vortices; red dots show vortices that have moved in the previous time step, black (fading to gray) tracers show the positions of the moving vortices for the 20 previous time steps.
Bottom panel: container angular velocity $\Omega_C$ versus time $t$.
Snapshots from the top five panels are marked with blue crosses. 
}
\label{fig:avalanche}
\end{figure}

Figure \ref{fig:avalanche} depicts the motion of the vortices during the glitch that begins at $t \approx 3080 T_0$, the largest glitch in the simulation. 
Overall the event is avalanche-like; it is triggered at a single location and spreads away in a wedge-like channel, as vortices knock-on their neighbours.
We show the progression of the avalanche in five stages in the top five panels.
The bottom panel shows the associated evolution of $\Omega_C$ during the avalanche, with each snapshot marked by a blue cross.
The red dots in the top five panels mark the vortices that are moving at that instant.
The progressively fainter black/grey circles indicate their positions at the previous 20 time steps.
Light grey dots show the positions of the vortices that have not moved further than $a$ in the previous 20 time steps.
Upon studying figure \ref{fig:avalanche}, we observe the following features.
\begin{itemize}
\item In the top panel, at $t/T_0 = 3079$, two nearby vortices toward the centre left unpin and begin moving radially outward.
Their motion does not have a noticeable effect on $\Omega_C$, which decreases steadily. 
\item In the second panel, at $t/T_0=3086$, more vortices are on the move. 
Some appear to unpin by direct proximity knock-on \citep{Warszawski2012};
others unpin by themselves shortly after a vortex closer to the boundary unpins and moves outwards.
Both these unpinning modes are observed in quantum mechanical, Gross-Pitaevskii simulations of smaller systems
\citep{Warszawski2011,Warszawski2012,Melatos2015}.
The collective outward motion tends to reverse the steady spin down.
Two other vortices unpin in another region of the container, towards the top right. 
They are likely unconnected to the avalanche in progress.
\item The third panel, at $t/T_0=3093$, shows many of the unpinned vortices leaving the container.
As vortices approach the boundary, they acquire a significant counter-clockwise azimuthal velocity, due to the presence of their corresponding image vortex across the boundary.
In contrast, unpinned vortices closer to the centre tend to move radially.
In this panel the two other unpinned vortices in the top right re-pin close to where they first unpinned.
\item The fourth panel, at $t/T_0=3104.5$, shows the vortex array after the majority of the unpinned vortices leave the container or re-pin.
Vortices move in a slow, inward, clockwise spiral before re-pinning in the evacuated region near where the avalanche begins.
Since the motion of the unpinned vortices is mostly azimuthal, rather than radial, spin up stops and $\Omega_C$ flattens out.
\item In the fifth panel, at $t/T_0 = 3116$ almost all of the vortices have re-pinned and $\Omega_C$ resumes steady spin-down.
During the avalanche, 68 vortices of the 1919 in the container at the beginning of the avalanche move an average distance of $\approx 0.18R$.
\end{itemize}

An important characteristic of avalanche dynamics is that avalanches relieve only a small amount of the accumulated stress, and the system remains in a marginal state close to the avalanche threshold in many places even immediately after an avalanche. 
To test this, we note that vortices are stationary when the pinning velocity vector $\mathbf{v}_{\rm pin}$ is exactly equal and opposite to the vorticity-induced velocity vector (from real and image vortices) $\mathbf{v}_{\rm induced}$.
With a Gaussian pinning potential, one has 
\begin{equation}
\mathbf{v}_{\rm pin} = V_0 \exp\left[- \frac{(\mathbf{x}-\mathbf{x}_k)^2}{2\xi^2}\right]
(\mathbf{x}-\mathbf{x}_k) \times \mathbf{e}_z
\label{eq:pinning velocity} \quad ,
\end{equation}
where $\mathbf{x}_k$ is the position of the $k$-th pinning centre and $\mathbf{e}_z$ is the unit vector in the z-direction. 
The pinning speed peaks at 
max$\vert \mathbf{v}_{\rm pin} \vert = V_0 \xi e^{-1/2}$ 
at
$\vert \mathbf{x}-\mathbf{x}_k \vert = \xi$.
A vortex leaves the region $\vert \textbf{x}-\textbf{x}_k \vert \leq \xi$ permanently if the absolute value of the sum of the first three terms in equations \eqref{eq:x velocity}--\eqref{eq:y velocity} exceeds max$\vert \mathbf{v}_{\rm pin} \vert$.
We can therefore characterise the marginal stability approximately by looking at the spatial distribution of 
$v_{\rm stress} = \vert \mathbf{v}_{\rm induced} - \boldsymbol{\Omega}_C \times \mathbf{x} \vert / \rm{max} \vert \mathbf{v}_{\rm pin} \vert$.

\begin{figure}
\includegraphics[scale=0.4]{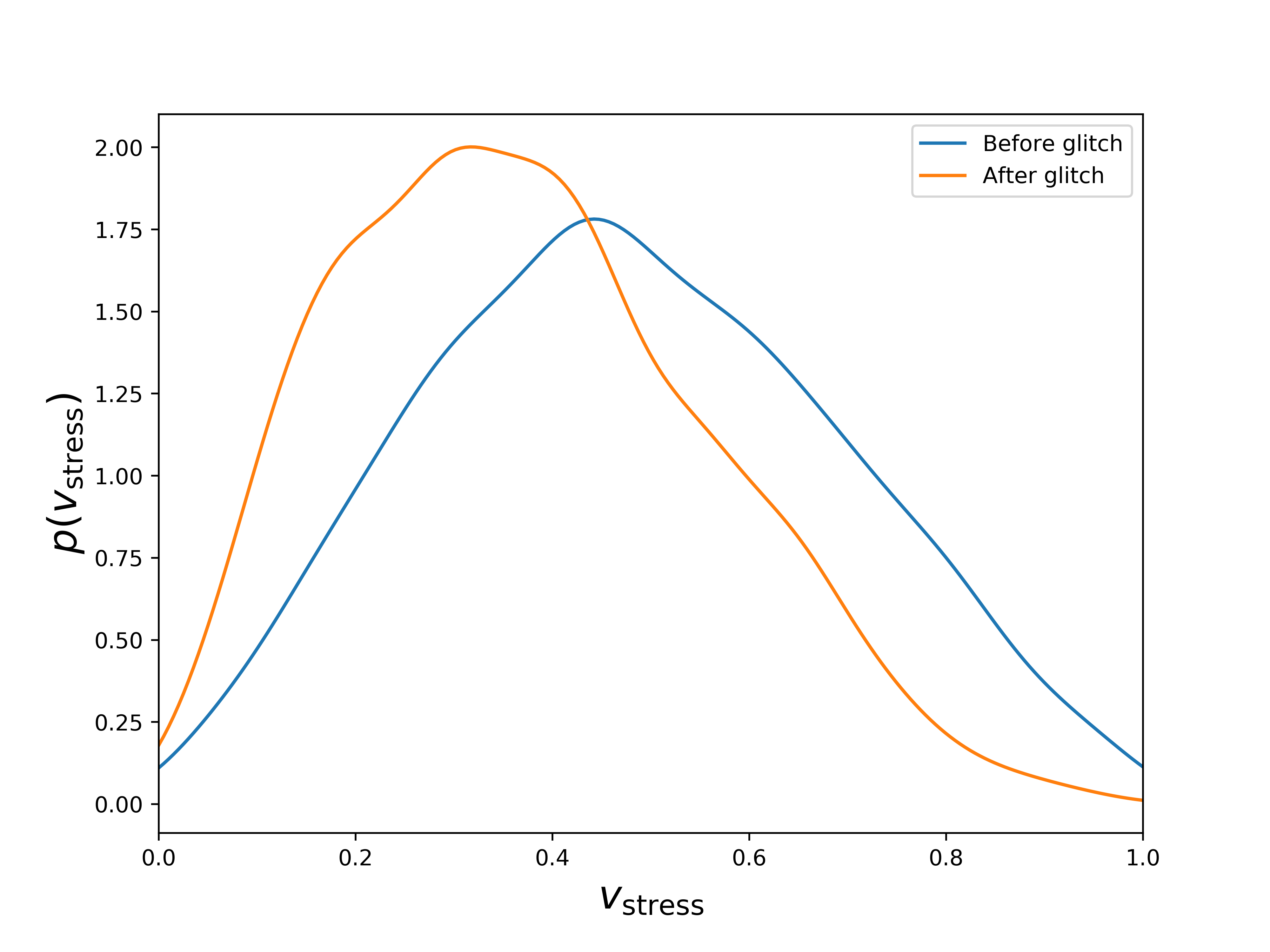}
\includegraphics[scale=0.45]{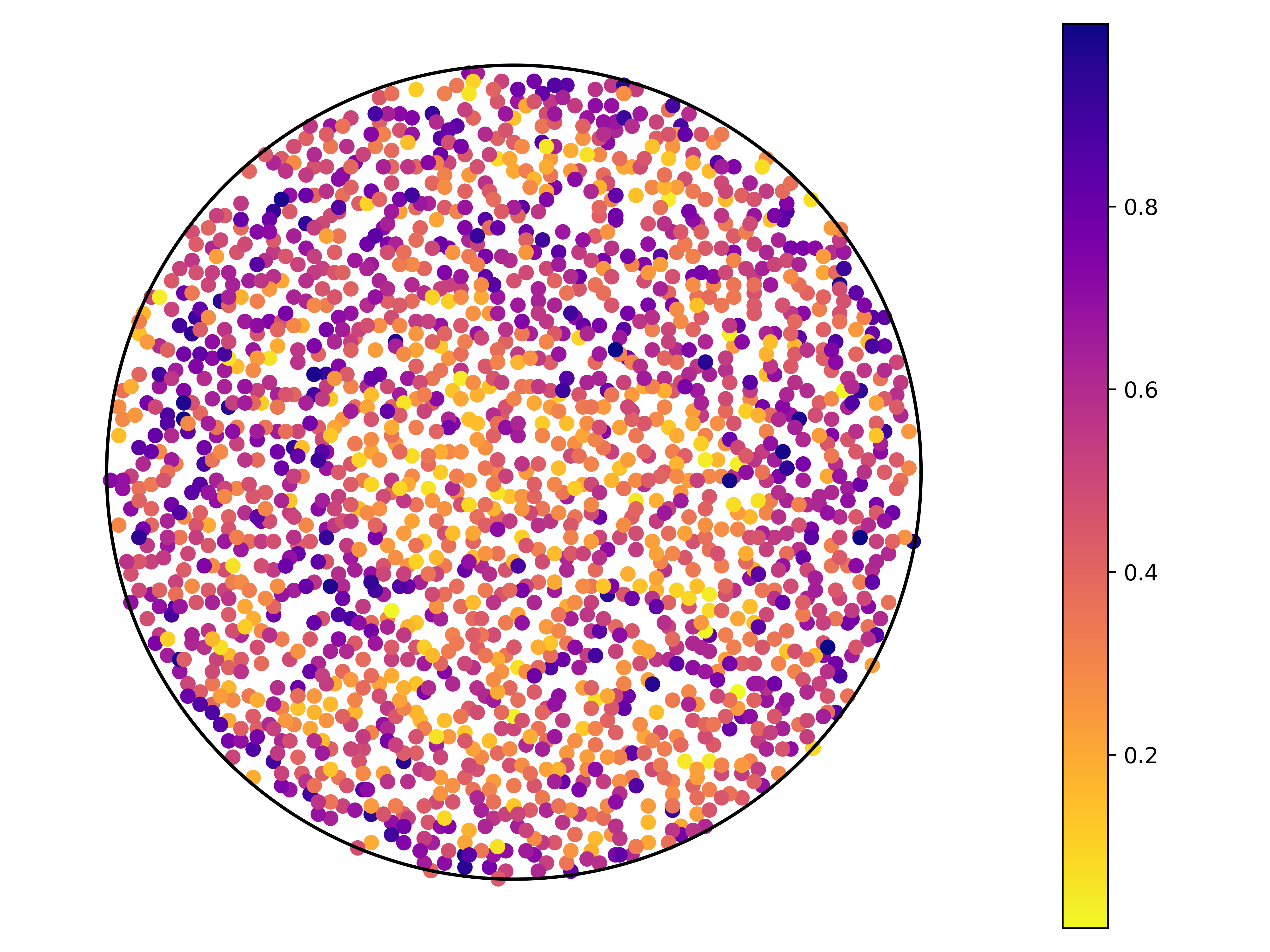}\\
\includegraphics[scale=0.45]{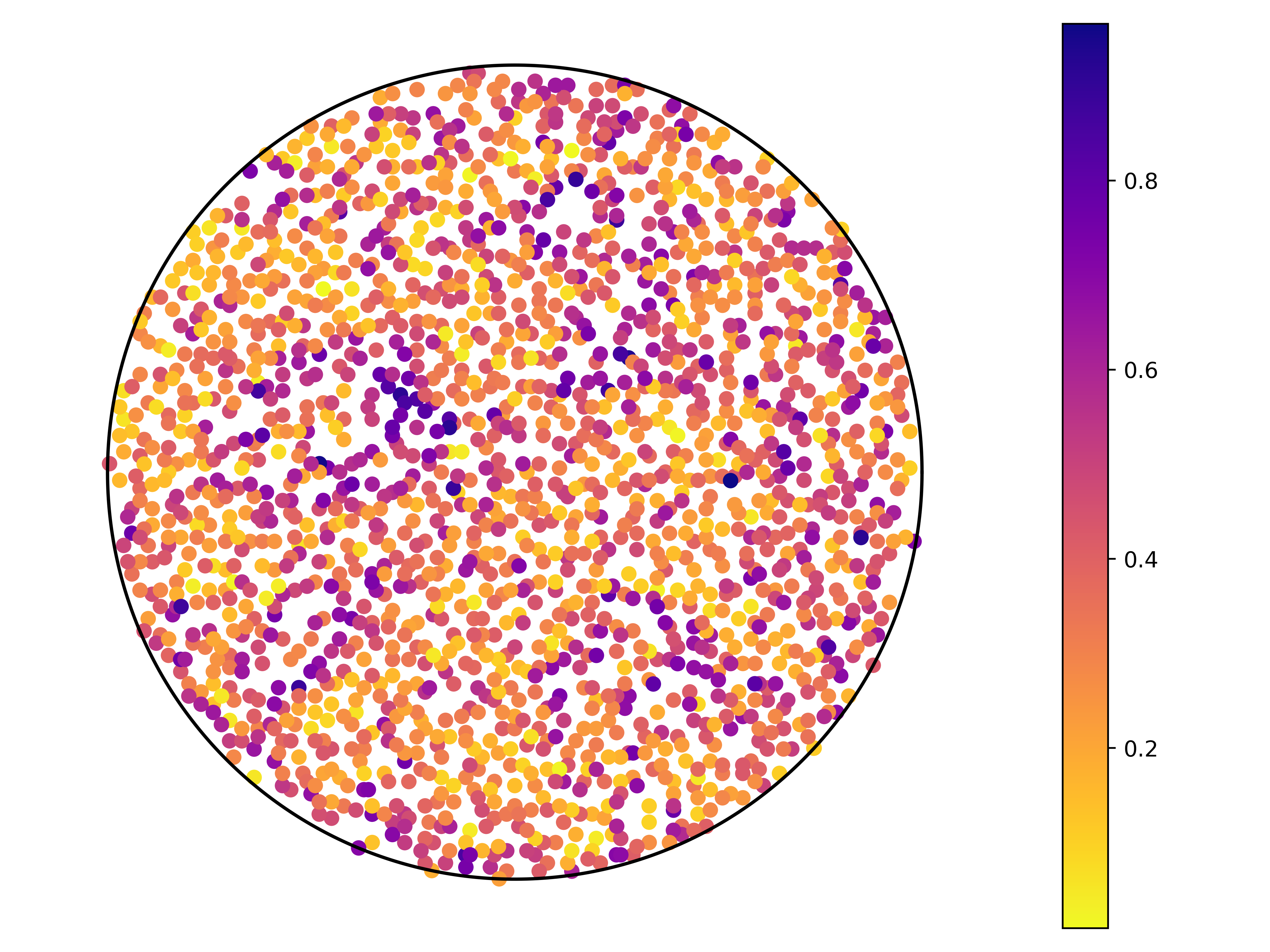}\\
\caption{Top panel: PDF of stress before (blue curve) and after (orange curve) the glitch, smoothed using a kernel density estimator.
Effective stress, parametrized by the speed $v_{\rm stress}$ before (middle panel) and after (bottom panel) a vortex avalanche.
Red dots indicate vortices with $v_{\rm stress} > 0.5$.}
\label{fig:stress}
\end{figure}

In Figure \ref{fig:stress}, we show the probability distribution function (PDF) of $v_{\rm stress}$ before and after the glitch shown in Figure \ref{fig:avalanche}.
We also show scatter plots of the vortex positions before and after the glitch, where vortices with $v_{\rm stress} > 0.5$ are coloured in red.
Figure \ref{fig:stress} conveys three important results.
Firstly, stress does not accumulate exclusively in the region where the avalanche occurs.
Stressed vortices occur throughout the container.
Secondly, while some stress is relieved by the glitch -- the median value of $v_{\rm stress}$ decreases from 0.47 to 0.36, and the full-width half-maximum of the PDF decreases from 0.57 to 0.50 -- a lot remains afterwards.
Thirdly, the remaining stress is not localised to the region outside the avalanche. Stressed vortices are more-or-less uniformly distributed throughout the container before and after the glitch, including where the avalanche occurred.

\subsection{Sizes and waiting times}
\label{subsec:size and waiting times}

\begin{figure}
\includegraphics[scale=0.6]{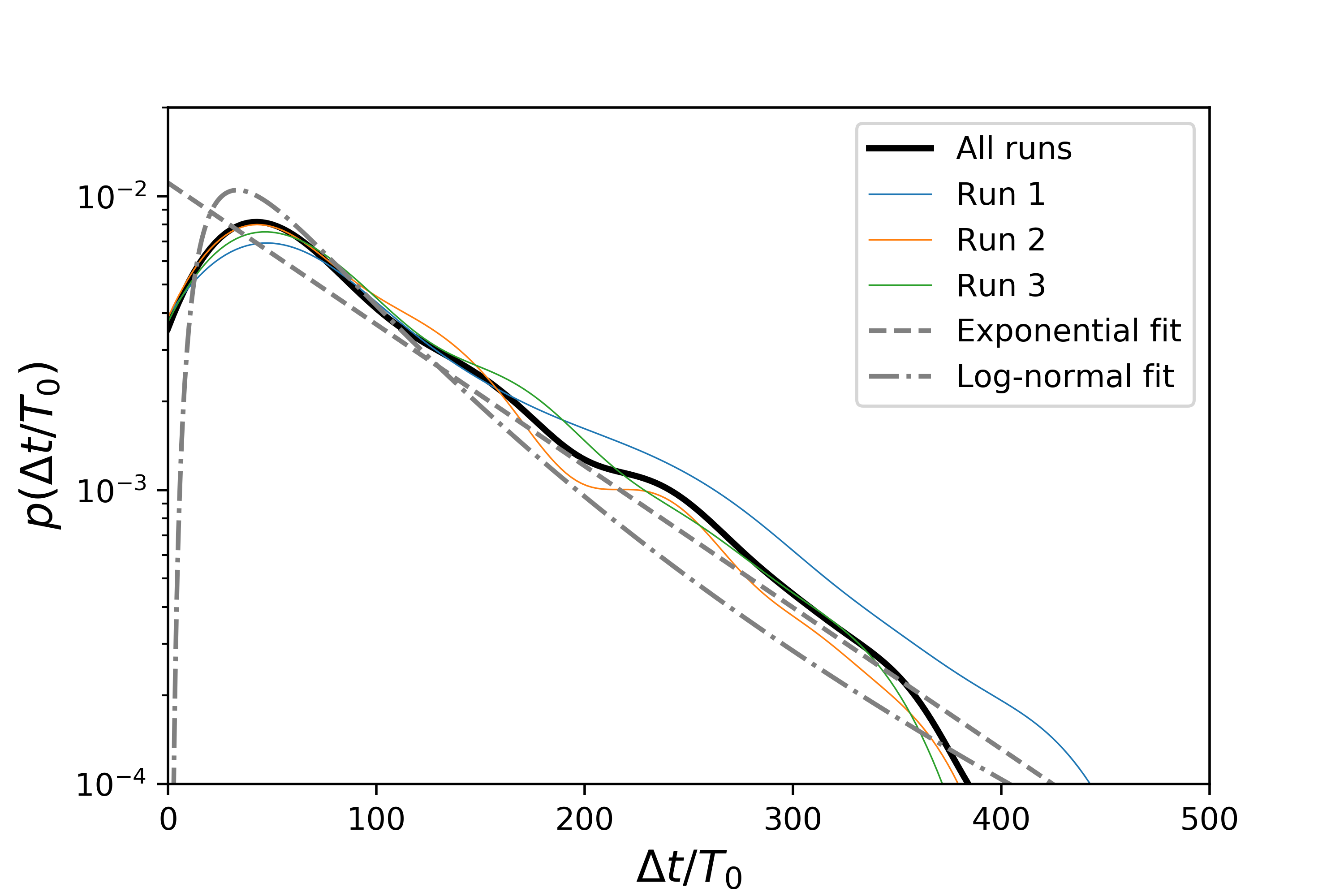}
\caption{Kernel density estimates of the waiting time PDF from three simulations with the default parameters in section \ref{subsec:simulation_description}. 
Black curve: aggregated data. 
Coloured curves: individual simulations.
Grey, dashed line: maximum likelihood exponential fit.
Grey, dot-dashed curve: maximum likelihood log-normal fit.} 
\label{fig:waiting time distributions}
\end{figure}

\begin{figure}
\includegraphics[scale=0.6]{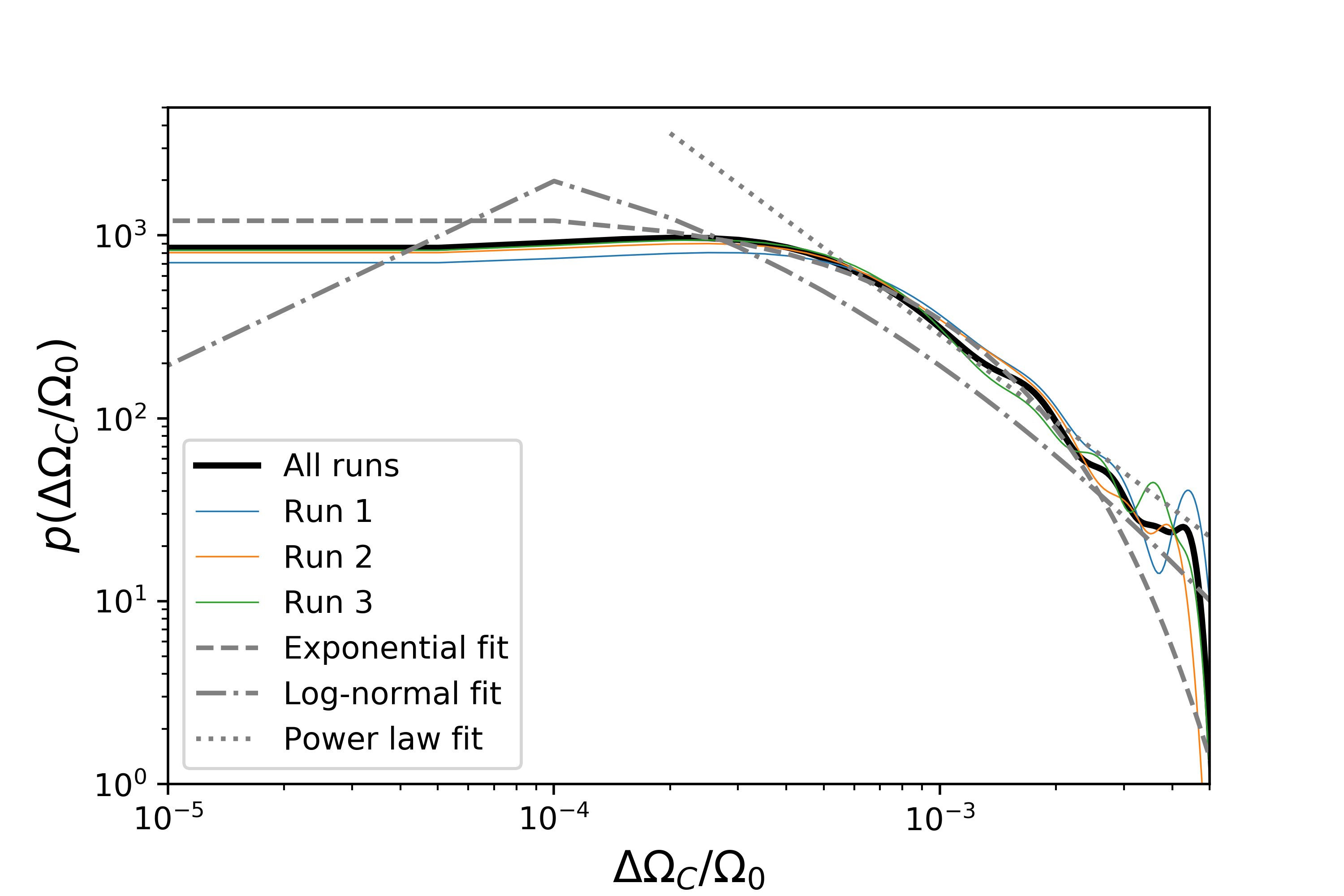}
\caption{Kernel density estimates of the size PDF from three simulations with the default parameters in section \ref{subsec:simulation_description}. 
Black curve: aggregated data. 
Coloured curves: individual simulations.
Grey, dashed curve: maximum likelihood exponential fit.
Grey, dot-dashed curve: maximum likelihood log-normal fit.
Grey, dotted line: power law fit for $\Delta \Omega_C / \Omega_0 > 2 \times 10^{-4}$.}
\label{fig:size distributions}
\end{figure}

Here we examine the statistical properties of glitch sizes and waiting times in the simulations.
We aggregate 314 glitches from three identically-initialized simulations with the default input parameters in section \ref{subsec:simulation_description} and calculate kernel density estimates of the waiting time and size PDFs.
Figures \ref{fig:waiting time distributions} and \ref{fig:size distributions} display the results for the waiting time and size PDFs respectively for both the aggregated data and the three individual simulations.
We also include fits to an exponential distribution, 
$p(x) = \lambda \exp(-\lambda x)$, 
and a log-normal distribution,
$p(x) = (2 \pi x^2 \sigma^2)^{-1/2} 
\exp [ - (\ln x - \mu)^2 / 2\sigma^2 ]$, where the fit parameters take their maximum likelihood values 
$\lambda = 1/\langle x \rangle$, 
$\mu = \langle \ln x \rangle$, and
$\sigma^2 = \langle(\ln x - \mu)^2 \rangle$
\citep{Howitt2018,Fuentes2019}.
For the size distribution, we also include a fit to a power law distribution, $p(x) = A x^{-a}$.
The fitting tool we use for the power law, \texttt{optimize.curve\_fit} in the \texttt{scipy} package, is unable to produce a fit for the entire data set.
However, if we consider just the tail of the data with $\Delta \Omega_C / \Omega_0 > 2 \times 10^{-4}$ (comprising $\approx 2/3$ of the glitches), we find a best-fit value of $a = 1.6$ for the power law index.

We find that both the waiting time and size distributions are well described by exponential and log-normal PDFs.
An Anderson-Darling test finds consistency for both sizes and waiting times with both exponential and log-normal distributions at the 99\% confidence level.
Similar studies on pulsar glitch data 
\citep{Melatos2008,Howitt2018}
and quantum mechanical Gross-Pitaevskii simulations of vortex avalanches
\citep{Warszawski2011}
often produce exponential waiting time PDFs and power-law size PDFs.
However, the observational data offer support for other functional forms too [eg. log-normal, Gaussian 
\citep{Fuentes2019} and are based on small samples ($\lesssim 50$ events per pulsar)].
The Gross-Pitaevskii simulations are also based on small samples ($\approx 10^2$ vortices) and include acoustic knock-on
\citep{Warszawski2012}, a process not present in this paper, which allows long-range unpinning.
In contrast, the simulations in this paper involve more reliable statistics, with $2 \times 10^3$ vortices, and $\approx 10^2$ glitches per simulation.
The applicability of the power-law PDF to only the larger glitches is interesting.
A similar result was shown in observational data by 
\citet{Janssen2006},
who suggested that it may be due to a population of undetected small glitches.
This may also be the case in our simulations. 
Slow-moving avalanches of few vortices may reduce the magnitude of the spin-down rate but without changing its sign.
These avalanches are not picked up by our glitch-finding algorithm.sign, and so not be picked up by our glitch-finding algorithm.

\subsection{Model variations}
\label{subsec:variations}

We do not attempt to fit the model in this paper to astrophysical data, because the values taken by the input parameters in a neutron star are uncertain.
It is useful, however, to develop a rough sense of how the avalanche dynamics depend on the input parameters.
We vary four quantities: the strength of pinning, parametrised by $V_0$, the density of pinning sites, parametrised by $a$, the strength of dissipation, parameterised by $\phi$, and the spin-down rate, parametrised by $N_{\rm ext}$.
We vary each parameter individually below and above the default value in section \ref{subsec:simulation_description}.
As well as the default values, we run simulations with $V_0=500$ and $V_0 = 2000$; $a = 0.025R$ and $a=0.005R$ (corresponding to a ratio of pinning sites/vortices of $\approx 1$ and $\approx 100$ respectively); $\phi = 0.01$ rad and $\phi = 0.5$ rad; and $N_{\rm ext}=5 \times 10^{-4}$ and $N_{\rm ext} = 2 \times 10^{-3}$. 
We perform an ensemble of three 2000-vortex simulations for each model variation and compute the total number of glitches, the mean waiting time between glitches, and the mean glitch size.

Table \ref{table:model_variations} shows a comparison of these quantities to the default simulation; cf. Table 9 in \citet{Warszawski2011}.
\begin{table}
	\begin{tabular}{|c|c|c|}
		\hline 
		Parameter & $\langle \Delta t/T_0 \rangle$ &  $\langle \Delta \Omega_C \rangle$  \\ 
		\hline 
		$V_0$ & $+$ &  $+$  \\ 
		$a$ & $+$ & .   \\ 
		$\phi$ & . &  .  \\ 
		$N_{\rm ext}$ & $-$ & .   \\ 
		\hline 
	\end{tabular} 
	\caption{Effect of varying simulation parameters (first column) on vortex avalanche statistics (columns two and three): mean waiting time between glitches $\langle \Delta t/T_0 \rangle$, and mean glitch size $\langle \Delta \Omega_C \rangle$.
	Symbols $+$ ($-$) indicate that $\langle \Delta t \rangle$ or $\langle \Delta \Omega \rangle$ increases (decreases) relative to the default value in section \ref{subsec:simulation_description} as the parameter in column one increases;
	``." indicates no noticeable effect.
	}
	\label{table:model_variations}
\end{table}
Plus (minus) signs in Table \ref{table:model_variations} indicate that the observable in each column increases (decreases) as the parameter increases.
Dots indicate no consistent change in the observable as the parameter varies.
We find that the average waiting time and glitch size both increase as the pinning strength, $V_0$, increases.
Increasing the spacing between pinning sites, $a$, (i.e. lowering the density of pinning sites) causes the average waiting time to increase and does not change the average glitch size.
Changing the drag angle, $\phi$, does not consistently push the average waiting time or size in one particular direction. In fact, $\langle \Delta t \rangle$ is higher in the default case ($\phi=0.1$) than its value with both $\phi = 0.01$ and $\phi=0.5$. 
Increasing the spin-down torque, $N_{\rm ext}$, results in more frequent glitches but does not change the average size.
Our results extend Table 9 in 
\citet{Warszawski2011}
by including $\phi$.
Where we can make direct comparisons, however, all of our results agree with 
\citet{Warszawski2011}.

The results in Table \ref{table:model_variations} are interpreted physically as follows.
Increasing the spin-down torque, $N_{\rm ext}$, does not appear to change the dynamics of the glitches.
It simply increases their frequency due to the more rapid build-up of stress.
As $V_0$ increases, vortices withstand more stress before unpinning and hence travel further before re-pinning, leading to more knock-on and hence larger glitches.
Since larger glitches release more stress, they occur less often.
This also explains the longer waiting times with stronger pinning; it takes longer to build up to the critical lag threshold after a large glitch.
As we increase the spacing between pinning sites, vortices are less likely to re-pin after unpinning. 
This leads to steady outward flow of vortices, rather than sudden, collective motion as the container decelerates, though some small avalanches still occur.
Because our glitch-finding algorithm counts only unpinning events that lead to a spin-up of the container, the steady flow is not picked up, leading to a reduced number of glitches with greater waiting times between them.

\begin{figure}
	\begin{center}
		\includegraphics[scale=0.19]{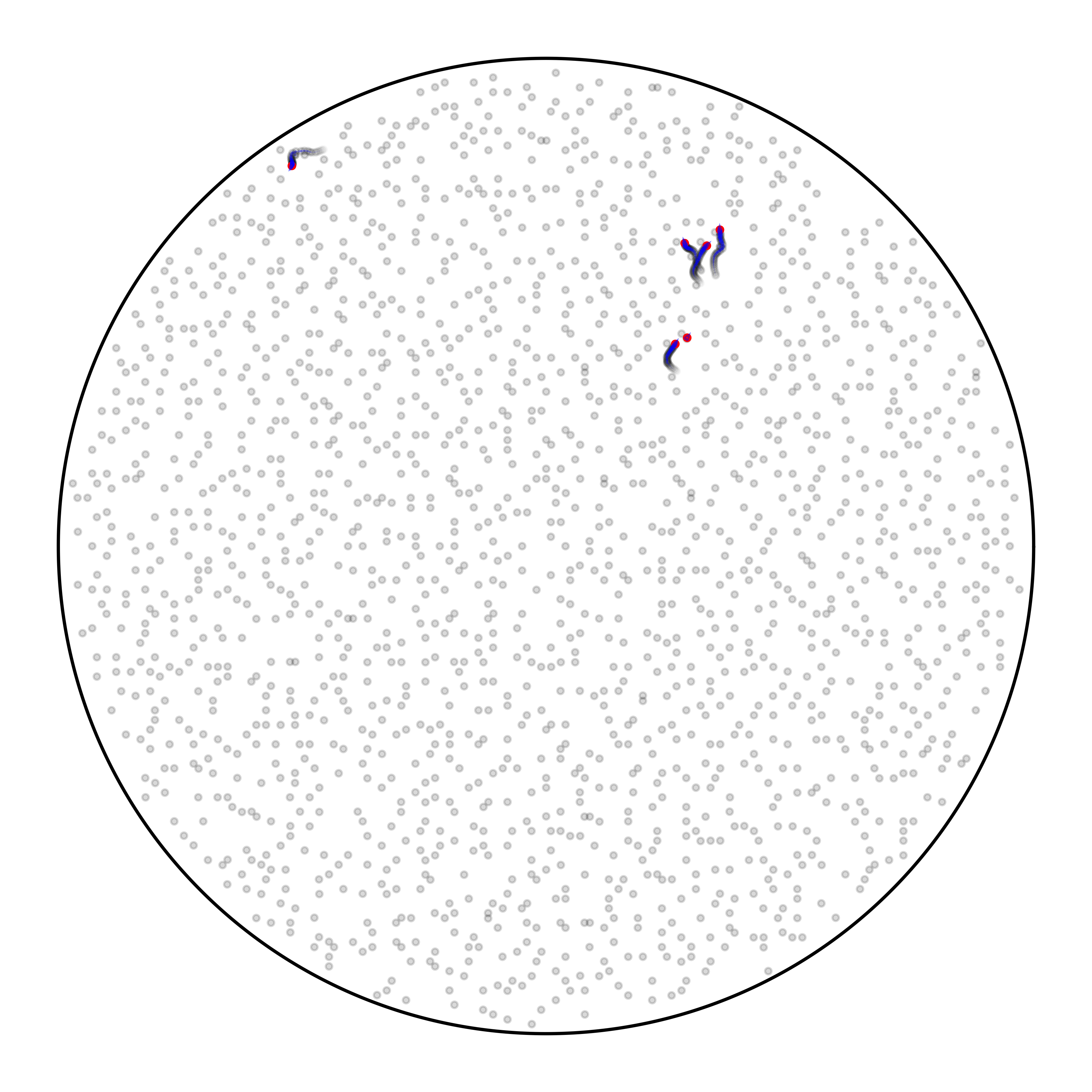}\\
		\includegraphics[scale=0.19]{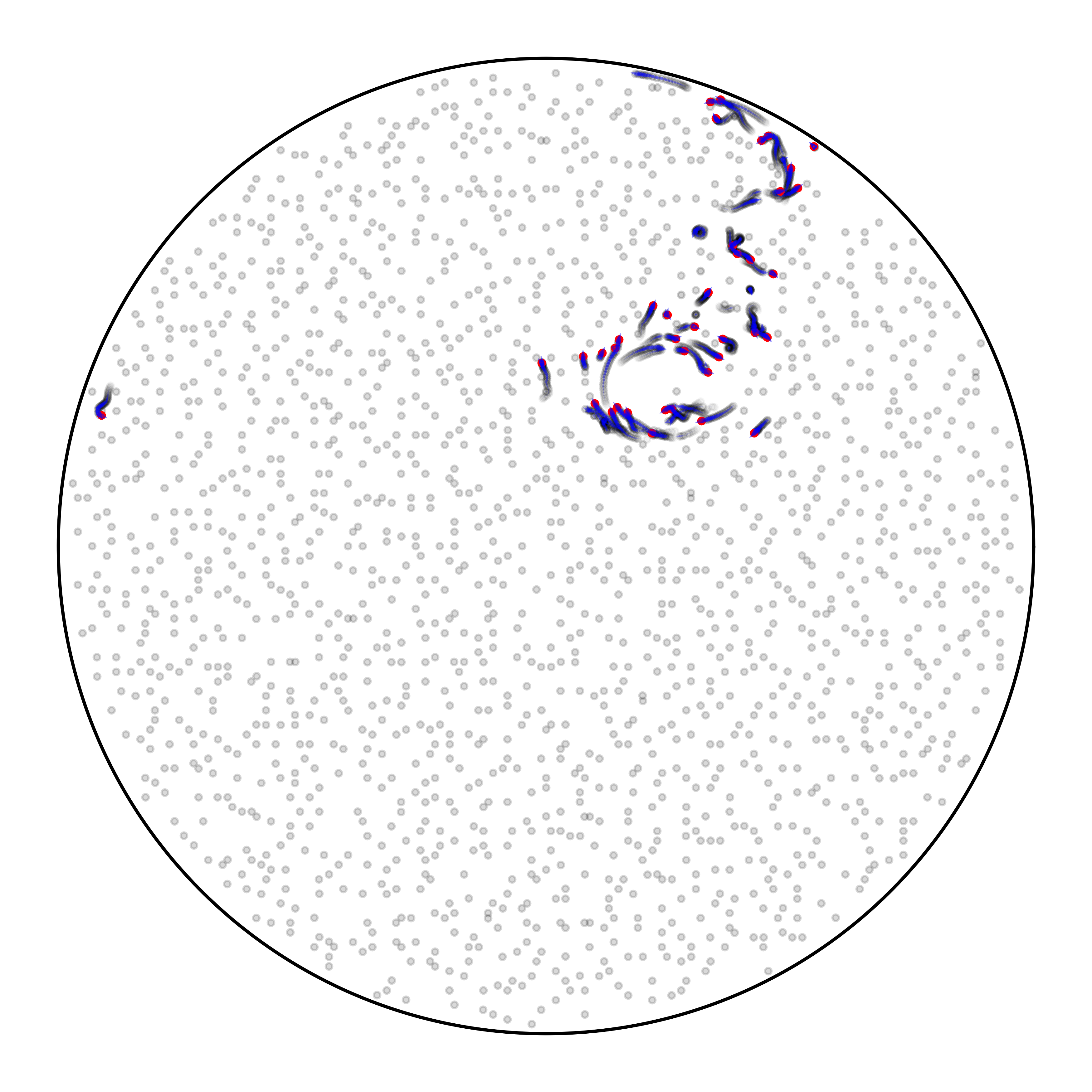}\\
		\includegraphics[scale=0.19]{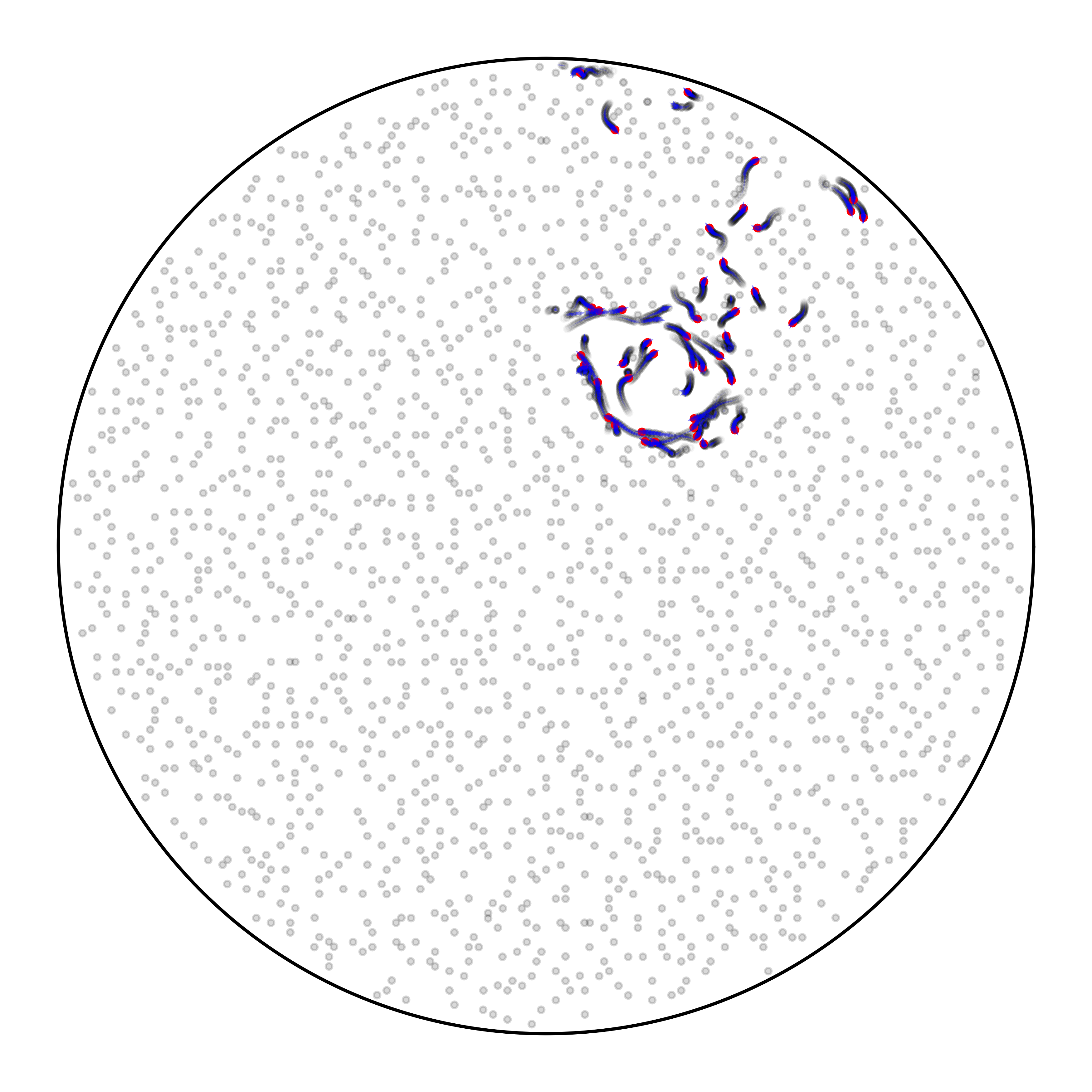}\\
		\includegraphics[scale=0.19]{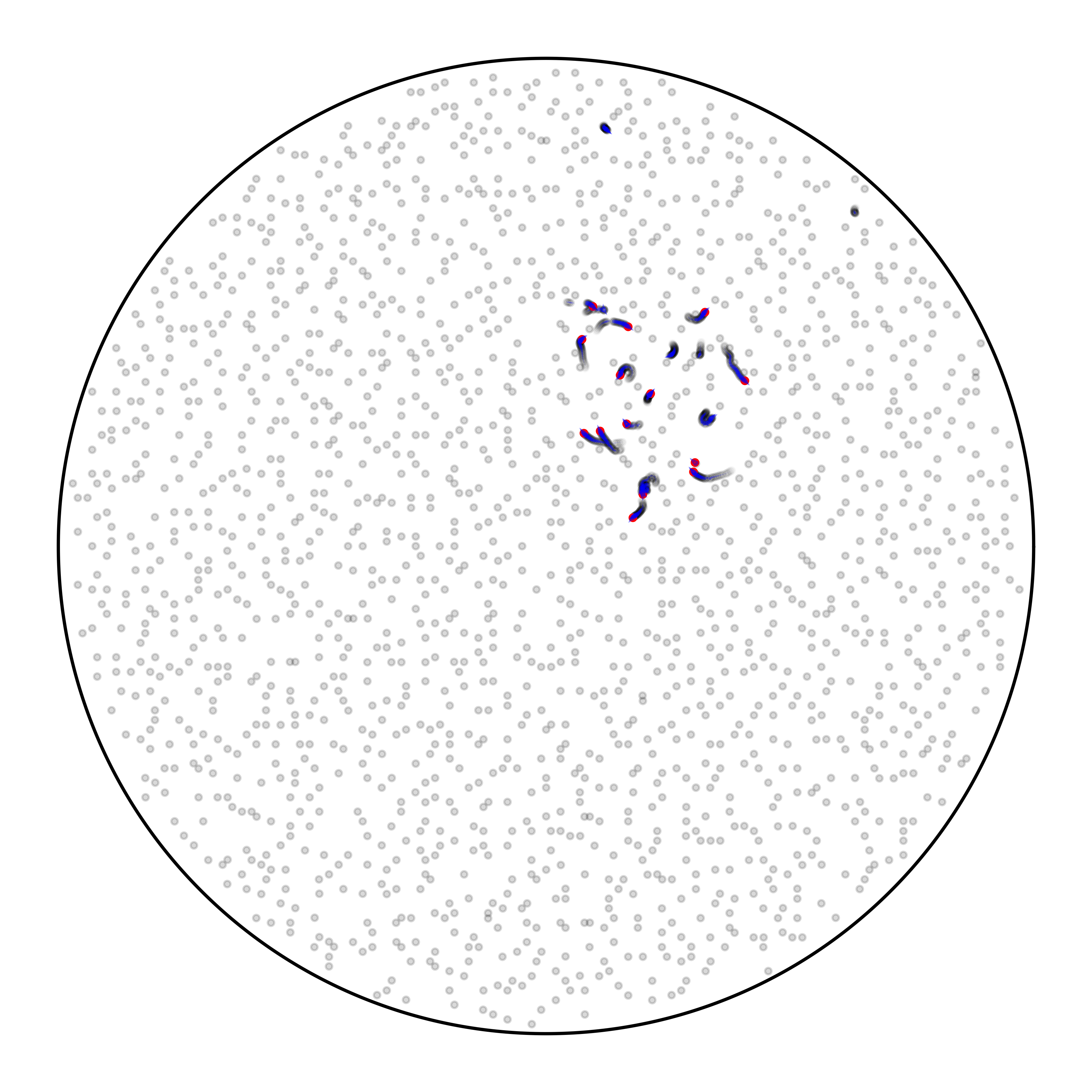}\\
		\includegraphics[scale=0.4]{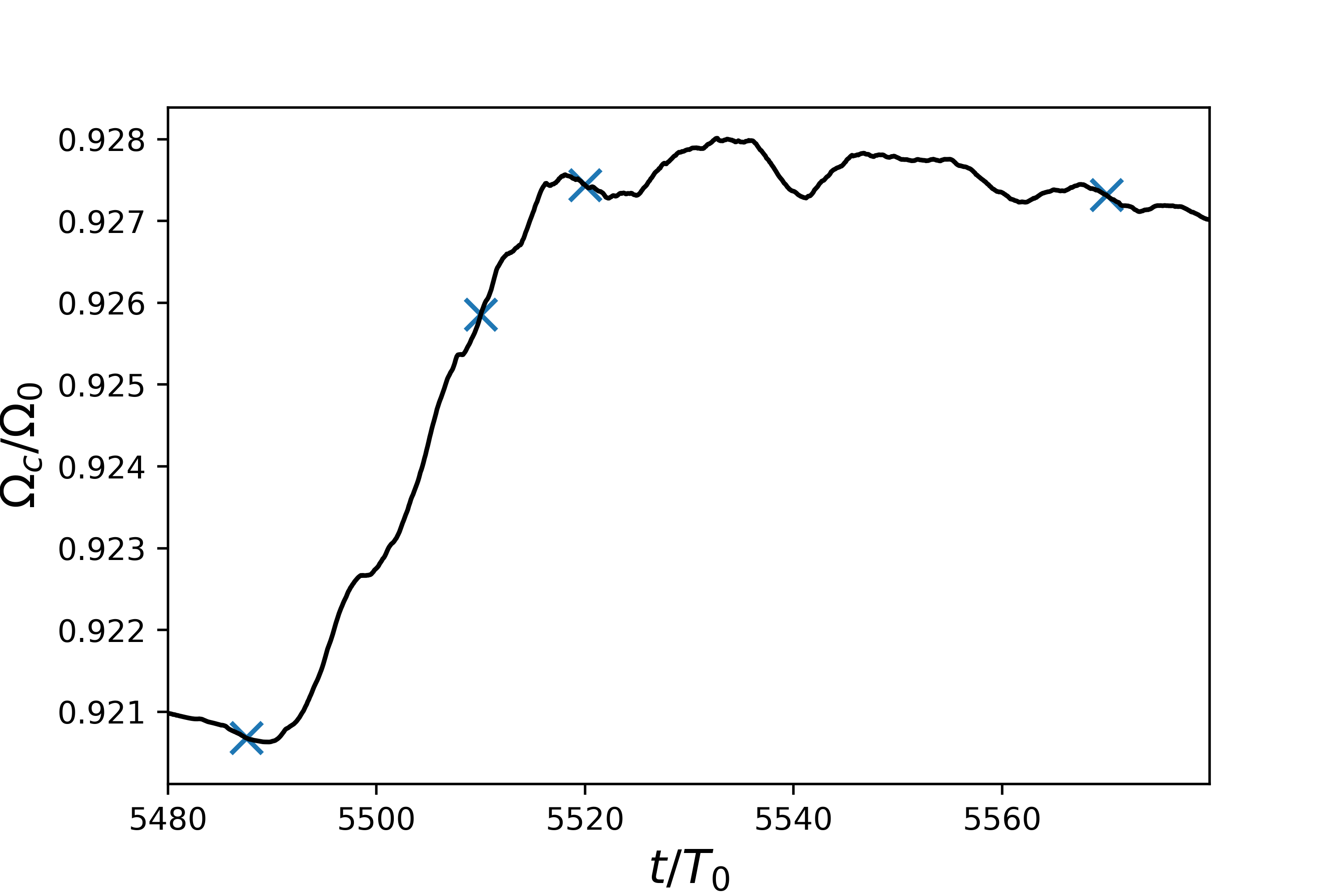}
	\end{center}
	\caption{Vortex motion during an avalanche in the weak dissipation simulation.		
	Top four panels are snapshots at (top to bottom) $t/T_0 =$ 5487.5, 5510, 5520, 5570.
	Grey dots show stationary vortices; red dots show vortices that have moved in the previous time step, black (fading to gray) tracers show the positions of the moving vortices for the 20 previous time steps.
	Bottom panel: container angular velocity $\Omega_C$ versus time $t$.
	Snapshots from the top four panels are marked with blue crosses. 
	}
\label{fig:pinball avalanche}
\end{figure}

The effect of changing the drag angle $\phi$ is complicated.
When vortices unpin, the circulatory motion induced by other vortices and pinning sites causes them to follow a spiral trajectory as they move outward.
Increasing the drag angle $\phi$ makes the trajectory more radial, leading to less knock-on and hence smaller glitches.
In contrast, when $\phi$ is lower, there is more knock-on, but unpinned vortices are more influenced by interactions with nearby vortices than by the dissipative radial motion.
They ``pinball'' throughout the vortex array, with some eventually re-pinning closer to the center than where they unpin. 
While the overall tendency is that vortices migrate outwards, it is a more protracted process than in the default case.
To better illustrate this effect, we examine the largest glitch in the weak dissipation simulation at high time resolution, cf. Figure \ref{fig:avalanche}, in Figure
\ref{fig:pinball avalanche}.
The top panel in Figure \ref{fig:pinball avalanche} shows the first few vortices that unpin at $t = 5487.5T_0$ and begin to move toward the edge of the container.
In the second panel, at $t = 5510 T_0$, more vortices unpin.
On average, vortices move radially outwards, but a significant number are also moving either inward or azimuthally.
In the third panel, at $t = 5520 T_0$, many vortices remain unpinned but the `swirling' motion dominates, so the container stops spinning up (see third blue cross on bottom panel).
In the fourth panel, at $t = 5570 T_0$, some vortices remain unpinned some time after the container ceases spinning up.
Their motion is disordered, leading to bumpy evolution of the spin frequency.

\subsection{Size-waiting time  correlations}
\label{subsec:correlations}

Auto- and cross correlations between glitch sizes and waiting times are an important diagnostic of the balance between crustal spin down and vortex unpinning
\citep{Fulgenzi2017,Melatos2018,Carlin2019a,Fuentes2019}.
Here we look for size and waiting time correlations in our simulations across all model variations.

Firstly, we look at cross correlations.
Two types of correlations are considered: forward correlations between size and the waiting time to the following glitch; and backwards correlations between size and the waiting time since the preceding glitch.
In general, such correlations are not expected in systems with SOC, and are not seen in Gross-Pitaevskii simulations of glitches
\citep{Warszawski2011}. 
Only one pulsar, PSR J0537$-$6910 has a strong forward correlation
\citep{Middleditch2006,Ferdman2018,Antonopoulou2018}. 
Several other pulsars appear to have weak forward correlations, though these are more uncertain due to the low number of observed glitches.
No backwards correlations are observed in the pulsar population
\citep{Melatos2018}.
The existence of size-waiting time cross correlations (or lack thereof) has been proposed as a means of falsifying meta-models of pulsar glitches as either a state-dependent Poisson process 
\citep{Fulgenzi2017,Melatos2018,Carlin2019a} 
or a Brownian stress-accumulation process 
\citep{Carlin2020}.
Both the Brownian and state-dependent Poisson meta-models predict positive forward correlations in certain regimes.

In Table \ref{tab:cross correlation table} we show the values of the Pearson correlation coefficients $r$ and the Spearman rank coefficients $\rho$ for both the forward and backward size-waiting time correlations for each of our model variations.
To assess significance, uncertainty is roughly the standard error $\sim r_{\pm} (N_{\rm g} -2)^{-1/2} \lesssim 0.1 r_{\pm}$.
\begin{table}
	\begin{tabular}{|c|l|l|l|l|l|}
		\hline
		Model variation & $N_g$ & $r_+$ & $r_-$ & $\rho_+$ & $\rho_-$ \\
		\hline
		Default & 311 & 0.35 & 0.04 & 0.44 & 0.03 \\
		Weak pinning & 360 & 0.25 & 0.07 & 0.24 & 0.03 \\
		Strong pinning & 204 & 0.33 & 0.004 & 0.37 & 0.04 \\
		Low density & 228 & 0.34 & -0.1 & 0.22 & -0.1 \\
		High density & 408 & 0.45 & 0.02 & 0.38 & 0.007 \\
		Weak dissipation & 392 & -0.12 & 0.14 & -0.19 & -0.01 \\
		Strong dissipation & 516 & 0.32 & -0.01 & 0.31 & -0.01\\
		Fast spin-down & 420 & 0.44 & -0.03 & 0.39 & -0.05 \\
		Slow spin-down & 188 & 0.36 & 0.1 & 0.34 & -0.06 \\
		\hline
	\end{tabular}
	\caption{Pearson correlation coefficients $r$ and Spearman rank coefficients $\rho$ for forward (+) and backward ($-$) size-waiting time correlations in each of our model variations.}
\label{tab:cross correlation table}
\end{table}
All of our model variations, with the exception of weak dissipation, show a weak positive forward correlation between glitch size and waiting time.
None show a significant backwards correlation.

In Table \ref{tab:autocorrelation table}, we show the auto-correlation in size and waiting times for all model variations.
\begin{table}
	\begin{tabular}{|c|c|c|c|c|}
		\hline
		& Waiting time & & Size \\
		Model variation & $\rho_{\Delta t}$ & p-value & $\rho_{\Delta \Omega}$ & p-value \\
		\hline
		Default & 0.068 & 0.23 & -0.037 & 0.52 \\
		Weak pinning & -0.024 & 0.65 & 0.0010  & 0.85 \\
		Strong pinning & 0.039 & 0.58 & -0.18 & 0.011 \\
		Low density & -0.21 & 0.0016 & 0.081 & 0.22 \\
		High density & 0.0045 & 0.93 & -0.012 & 0.81 \\
		Weak dissipation & -0.19 & 0.00012 & 0.10 & 0.041 \\
		Strong dissipation & -0.040 & 0.36 & 0.064 & 0.15 \\
		Fast spin-down & 0.0066 & 0.90 & -0.083 & 0.12 \\
		Slow spin-down & 0.024 & 0.75 & 0.071 & 0.33 \\
		\hline
	\end{tabular}
	\caption{Spearman autocorrelation coefficient $\rho$ and p-values for glitch waiting times $\Delta t$ and sizes $\Delta \Omega$ in each of our model variations [cf. Table 1 in \citet{Carlin2019a}].}
	\label{tab:autocorrelation table}
\end{table}
None of the model variations show a significant positive autocorrelation for size or waiting time.
The low density and weak dissipation cases have a statistically-significant (p-value $<$ 0.05) weak negative autocorrelation in waiting time and the strong pinning and weak dissipation cases have a statistically-significant weak negative autocorrelation in the size.
In the case of the weak dissipation size autocorrelation, the significance is marginal.
We perform 18 independent significance tests, so it is likely that at least one p-value is less than 0.05 even if the null hypothesis of no autocorrelation is correct in all cases.

The results in Tables \ref{tab:cross correlation table} and \ref{tab:autocorrelation table} are broadly consistent with both meta-models.
There are no strong cross-correlations or autocorrelations in any of our models, which is consistent with what is observed in self-organized critical systems 
\citep{Jensen1998} 
and Gross-Pitaevskii simulations 
\citep{Warszawski2011}, 
albeit counter-intuitive for a threshold-triggered stress-release process.
With respect to the state-dependent Poisson model, the cross-correlations suggest that we are in the fast-driving regime
[see figure 13 in \citet{Fulgenzi2017}]. 
This is not surprising. In order to produce statistically useful numbers of glitches, computational constraints require spin-down rates far greater than those observed in even the most rapidly decelerating pulsars.
The models with negative autocorrelations in waiting time and size are inconsistent with the Brownian meta-model, and are consistent with the state-dependent Poisson model only in a restricted subset of parameter space 
\citep{Carlin2019a,Carlin2020}.

\section{Conclusion}
\label{sec:conclusion}

Superfluid vortex avalanches have long been suggested as a mechanism for pulsar glitches, but the existence of avalanche behaviour even in simplified models of neutron stars has not been demonstrated in systems with more than $\approx$ 100 vortices.
We have written a two-dimensional $N$-body solver based on the vortex filament model, including dissipation. 
Our results exhibit avalanche behaviour across a wide range of physical parameters for large-scale systems with $N \lesssim 5 \times 10^3$ vortices, demonstrating that vortex avalanches are ubiquitous when vortices are pinned in a decelerating container.
Our results agree qualitatively with previous Gross-Pitaevskii simulations with $\approx 10^2$ vortices, despite the lack of acoustic knock-on in our classical point-vortex model.
Comparing our simulations to the observed population of pulsar glitches, we find similar waiting time distributions, but our size distributions differ. 
We find weak cross-correlations between glitch sizes and waiting times for almost all of our model variations, a finding consistent with meta-models of pulsar glitches as a state-dependent Poisson processes or a Brownian stress-accumulation process.
The correlation results are also qualitatively consistent with the observed population, in which statistically significant correlations have only been observed in one pulsar, PSR J$0537-6910$.
We see weak negative autocorrelations in some of our models, which are inconsistent with the Brownian motion meta-model and restrict the parameter of the state-dependent model.

Advances in glitch modelling and detection are paving the way for falsifying specific glitch mechanisms, such as superfluid vortex avalanches. 
On the modelling front, large scale $N$-body simulations like those in this paper make specific falsifiable predictions about the long-term statistics of glitches and their individual profiles in time.
Future improvements include relaxing the simplifying assumptions in the model, such as moving to three dimensions and including vortex tension 
\citep{Link2009}.
On the observational front, improvements include analyzing the completeness of existing datasets and enlarging the glitch sample with next generation pulsar observing campaigns with instruments such as the Square Kilometre Array. 
All of this can be combined with stress-release meta-models which make microphysics-agnostic predictions about the long-term glitch statistics, as exemplified by the discussion in section \ref{subsec:correlations} in this paper.

\section*{Acknowledgments}

We thank the referee, Andreas Reisenegger, for a careful reading of this paper and suggesting valuable improvements.
GH and AM acknowledge support from the the Australian Research Council (ARC) through the Centre of Excellence for Gravitational Wave Discovery (OzGrav) (grant number CE170100004) and an ARC Discovery Project (grant number DP170103625).
BH acknowledges support from the National Science Centre, Poland (NCN), via grant SONATA BIS 2015/18/E/ST9/00577.
GH acknowledges support from the University of Melbourne through a Melbourne Research Scholarship and a Faculty of Science Travelling Scholarship. 
Simulations in this paper were run on the Spartan HPC system at the University of Melbourne
\citep{spartan}.

\section*{Data availability}

Simulation data and code used in this paper can be made available upon request by emailing the corresponding author.

\bibliographystyle{mn2e}
\bibliography{vortex}

\label{lastpage}

\end{document}